\documentclass[draftcls,onecolumn]{IEEEtran}
\IEEEoverridecommandlockouts \ifx\pdfoutput\undefined
\usepackage{graphicx}
\else
\usepackage[pdftex]{graphicx}
\fi
\usepackage{cite, amsmath, amsfonts, amssymb, psfrag}
\newtheorem{proposition}{{Proposition}}
\newtheorem{definition}{{Definition}}
\newtheorem{theorem}{{Theorem}}
\newtheorem{lemma}{{Lemma}}
\newtheorem{example}{{Example}}
\newtheorem{corollary}{{Corollary}}
\newtheorem{remark}{{Remark}}
\newcommand {\hb}{\text{\boldmath{$h$}}}
\newcommand{\abu}{\underline{\text{\boldmath{$a$}}}}
\newcommand{\bbu}{\underline{\text{\boldmath{$b$}}}}
\newcommand {\xb}{\text{\boldmath{$x$}}}
\newcommand {\Xb}{\text{\boldmath{$X$}}}
\newcommand{\xbu}{\underline{\text{\boldmath{$x$}}}}

\newcommand {\yb}{\text{\boldmath{$y$}}}
\newcommand {\zb}{\text{\boldmath{$z$}}}
\newcommand {\wb}{\text{\boldmath{$w$}}}
\newcommand{\snr}{\mathsf{S\hspace{-0.011in}N\hspace{-0.011in}R}}
\newcommand{\sfP}{\mathsf{P}}
\DeclareMathAlphabet{\mathpzc}{OT1}{pzc}{m}{it}

\begin{document}

\title{Capacity Results for Block-Stationary Gaussian Fading Channels with a Peak Power Constraint}

\author{\authorblockN{Jun Chen and Venugopal V. Veeravalli}\\
\authorblockA{ECE Department  \& Coordinated Science Lab\\ University of Illinois at Urbana-Champaign\\
Urbana, IL 61801, USA\\
Email: junchen,vvv@uiuc.edu}
%\and
%\authorblockN{Jun Chen}
%\authorblockA{ECE Department  \& Coordinated Science Lab\\ University
%of Illinois at Urbana-Champaign\\
%Urbana, IL 61801, USA\\
%Email: junchen@ifp.uiuc.edu}
 }
% avoiding spaces at the end of the author lines is not a problem with
% conference papers because we don't use \thanks or \IEEEmembership
%Homer Simpson\authorrefmark{2},
%Montgomery Scott\authorrefmark{3} and
%Eldon Tyrell\authorrefmark{4}}
%Telephone: (800) 555--1212, Fax: (888) 555--1212}
%\authorblockA{\authorrefmark{4}Tyrell Inc., 123 Replicant Street, Los Angeles, California 90210--4321}}

\maketitle

\begin{abstract}
We consider a peak-power-limited single-antenna block-stationary
Gaussian fading channel where neither the transmitter nor the
receiver knows the channel state information, but both know the
channel statistics.  This model subsumes most previously studied Gaussian fading models. We first compute the asymptotic channel capacity in the high SNR regime and show that the behavior of channel
capacity depends critically on the channel model. For the special
case where the fading process is symbol-by-symbol stationary, we also reveal a
fundamental interplay between the codeword length, communication
rate, and decoding error probability. Specifically, we
show that the codeword length must scale with SNR in order to
guarantee that the communication rate can grow logarithmically
with SNR with bounded decoding error probability, and we find a
necessary condition for the growth rate of the codeword length. We
also derive an expression for the capacity per unit energy.
Furthermore, we show  that the capacity per unit energy is
achievable using temporal ON-OFF signaling with optimally
allocated ON symbols, where the optimal ON-symbol allocation scheme
may depend on the peak power constraint.

\end{abstract}


\section{Introduction}

The capacity analysis of {\em noncoherent} fading channels has received
considerable attention in recent years since it provides the
ultimate limit on the rate of reliable communication on such
channels.

Proposed approaches to modeling noncoherent fading
channels can be classified into two broad categories. The first is to model the fading process as a {\em
block-independent} process. In the standard version of this
model\cite{MH99}, the channel remains constant over blocks
consisting of $T$ symbol periods, and changes independently from
block to block. The second is to model the fading process as a
symbol-by-symbol {\em stationary} process. In this model, the
independence assumption is removed, but the block structure is not
allowed. Somewhat surprisingly, these two models lead to very
different capacity results. For the standard block fading model, the
capacity is shown \cite{MH99, ZT02} to grow logarithmically with
SNR, while for the symbol-by-symbol stationary model, the capacity
grows only double-logarithmically in SNR at high SNR if the fading
process is {\em regular} \cite{LM03,LM06,KL05}. For
symbol-by-symbol stationary Gaussian fading channels, if the
Lebesgue measure of the set of harmonics where the spectral
density of the fading process is zero is positive, the fading
process is {\em nonregular} and the capacity grows logarithmically
with SNR \cite{L05}.  This result is consistent with the capacity
result for block-independent fading channels in the sense that the
$\log\snr$ behavior in the high SNR regime results from the rank
deficiency of the correlation matrix of the fading process.
%VVV
This point was elucidated in \cite{LV04} where a {\em time-selective} block fading model was considered in which the rank of the correlation within the block is allowed to be any number between one and the blocklength.

However, the mechanisms that cause the rank deficiency in the
block-independent fading and nonregular symbol-by-symbol
stationary models are different. For the block-independent fading
model, the rank deficiency happens within each block. But for the
nonregular symbol-by-symbol stationary fading channel model, the
correlation matrix of the fading process over any finite block can
still be full-rank; the rank deficiency in this case is in the
asymptotic sense. In general, the rank deficiency of the
correlation matrix can be affected by both the short timescale
correlation of the fading process as in the block-independent
fading model and large timescale correlation as in the
symbol-by-symbol stationary channel model. In order to capture
both of these effects, we model the fading process as a {\em
block-stationary} Gaussian process.
%VVV
%

The block-stationary model was introduced and justified in
\cite{LV04}. We summarize the main points of the justification
here. In the block-independent fading model, the channel is
assumed to change in an i.i.d. manner from block to block. The
independence can be justified in certain time division or
frequency hopping systems where the blocks are separated
sufficiently in time or frequency to undergo independent fading.
The independence assumption is also convenient for
information-theoretic analysis as it allows us to focus on one
block in studying the capacity. If the blocks are not separated
far enough in time or frequency, the fading process can be
correlated across blocks and the block-stationary model is more
appropriate in this scenario. Without time or frequency hopping,
the channel variations from one block to the next are dictated by
the long term variations in the scattering environment. If we
assume that the variations in average channel power are
compensated for by other means such as power control, it is
reasonable to model the variation from block to block as
stationary and ergodic.
\begin{remark} \label{rem:blockvsstat}
The block-stationary model does not imply that the fading process
is stationary on a symbol-by-symbol basis as in the analysis of
\cite{LM03,L05}. But as  explained in \cite{LV04}, the
symbol-by-symbol stationary model is not realistic for time
intervals that are larger than that corresponding to a few
wavelengths. For this reason it may be more accurate to model the
fading process using a block fading model with possible
correlation across blocks than it is to model it as a
symbol-by-symbol stationary process. From the
viewpoint of analysis, the block-stationary model generalizes all previously
considered models discussed above and therefore so do the capacity
results for this model. More importantly, the block-stationary
model provides us with a framework to study the interplay between many
aspects of fading channels which are not captured in the
aforementioned models, and allows us to identify the properties
that are shared by different models and the properties that depend
on channel modelling.
\end{remark}

The channel capacity for the block-stationary model was only
studied in \cite{LV04} under certain constraints on the
correlation structure across blocks, which essentially disallow
rank deficiency over the large timescale. In this paper, we
conduct a more complete study of the capacity for this channel
model.

\section{Notation and System Model}
\subsection{Notation}
The following notation is used in paper. For deterministic
objects, uppercase letters denote matrices, lowercase letters
denote scalars, and underlined lowercase letters denote vectors.
Random objects are identified by corresponding boldfaced letters.
For example, $\Xb$ denotes a random matrix, $X$ denotes the
realization of $\Xb$, $\xbu$ denotes a random vector, and $\xb$
denotes a random scalar. For simplicity, sometimes we also use
$\xb^n$ to denote the random vector
$(\xb_1,\xb_2,\cdots,\xb_n)^\top$. Although uppercase letters are
typically used for matrices, there are some exceptions, and these
exceptions are noted explicitly in the paper. The operators
$\det$, $\mbox{tr}$, $*$, $\top$, and $\dagger$ denote
determinant, trace, conjugate, transpose and conjugate transpose,
respectively. We let $I_M$ denote the $M\times M$ identity matrix
for any positive integer $M$, and let $\mbox{var}(\abu|\bbu)$
denote
$\mathbb{E}[(\abu-\mathbb{E}(\abu|\bbu))(\abu-\mathbb{E}(\abu|\bbu))^\dagger]$
for random vectors $\abu$ and $\bbu$.
\subsection{System Model}
We consider a discrete-time channel whose time-$t$ complex-valued
output $\yb_t\in\mathbb{C}$ is given by
\begin{eqnarray}
\yb_t=\hb_t\xb_t+\zb_t\label{channelmodel}
\end{eqnarray}
where $\xb_t\in\mathbb{C}$ is the input at time $t$ with peak
power constraint $|\xb_t|^2\leq\snr$; $\{\hb_t\}$ models the
fading process; and $\{\zb_t\}$ models additive noise. The
processes $\{\hb_t\}$ and $\{\zb_t\}$ are assumed to be
independent and have a joint distribution that does not depend on
the input $\{\xb_t\}$. We assume that $\{\zb_t\}$ is a sequence of
i.i.d. circularly symmetric complex-Gaussian random variables of
zero mean and unit variance, i.e., $\zb_t\sim\mathcal{CN}(0,1)$.
We assume that the fading process $\{\hb_t\}$ is a
block-stationary process with $\hb_t\sim\mathcal{CN}(0,1)$ and
block length $T$, i.e.,
$\{\underline{\hb}_k=(\hb_{kT+1},\hb_{kT+2},\cdots,\hb_{kT+T})^\top\}_k$
is a vector-valued stationary process. Furthermore, we assume that
$\{\underline{\hb}_k\}$ is an ergodic process with a matrix spectral density function $S(e^{j\omega})$, $-\pi\leq\omega\leq\pi$.
Specifically,
\begin{eqnarray*}
S(e^{j\omega})=\sum\limits_{i=-\infty}^\infty R(i)e^{-j\omega i}
\end{eqnarray*}
where
$R(i)=\mathbb{E}\underline{\hb}_k\underline{\hb}^\dagger_{k-i}$.
Since $R(i)=R^\dagger(-i)$,  $i\in\mathbb{Z}$, it is not hard to
check that $S(e^{j\omega})$ is Hermitian, i.e.,
$S(e^{j\omega})=S^\dagger(e^{j\omega})$. Moreover, we have
$S(e^{j\omega})\succcurlyeq 0$ ($-\pi\leq\omega\leq\pi$), i.e.,
$S(e^{j\omega})$ is a positive semi-definite matrix.

There is an interesting relation between the matrix spectral
density function and the asymptotic prediction error.
Specifically, for the block stationary process $\left\{\hb_t+\frac{1}{\sqrt{\snr}}\zb_t\right\}$, define the following prediction error covariance matrices:
 \begin{eqnarray*}
\Sigma(\snr)&\triangleq&\mbox{var}\left(\left.\left(\hb_1+\frac{1}{\sqrt{\snr}}\zb_1,\hb_2+\frac{1}{\sqrt{\snr}}\zb_2,\cdots,\hb_T+\frac{1}{\sqrt{\snr}}\zb_T\right)^\top\right|\left\{\hb_t+\frac{1}{\sqrt{\snr}}\zb_t\right\}_{t=-\infty}^0\right),\\
\Sigma(\infty)&\triangleq&\mbox{var}\left(\left.\left(\hb_1,\hb_2,\cdots,\hb_T\right)^\top\right|\left\{\hb_t\right\}_{t=-\infty}^0\right).
\end{eqnarray*}
Then $\Sigma(\snr)$ and $\Sigma(\infty)$ are related to
the matrix spectral density function $S(e^{j\omega})$ of $\{\hb_t\}$ as \cite{R67}
\begin{eqnarray}
\det\left[\Sigma(\snr)\right]&=&\exp\left\{\frac{1}{2\pi}\int_{-\pi}^{\pi}\log\det\left[S(e^{j\omega})+\frac{1}{\snr}I_T\right]\mbox{d}\omega\right\},\label{identity2}\\
\det\left[\Sigma(\infty)\right]&=&\exp\left\{\frac{1}{2\pi}\int_{-\pi}^{\pi}\log\det\left[S(e^{j\omega})\right]\mbox{d}\omega\right\}.\label{identity1}
\end{eqnarray}

The remainder of this paper is organized as follows. In
Section~\ref{sec:asym_cap}, we establish single-letter upper and
lower bounds on channel capacity, and use these bounds to analyze
the asymptotic capacity in the high SNR regime. In Section
~\ref{sec:discussion}, we discuss the robustness of the asymptotic
capacity results, and the interplay between the codeword length,
communication rate and decoding error probability. In
Section~\ref{sec:cap_energy}, we adapt the formula of Verd\'{u}
for capacity per unit cost \cite{V90} to our channel model, and
use it to derive an expression for the capacity per unit energy in
the presence of a peak  power constraint. We summarize our results
in Section~\ref{sec:concl}.

\section{Asymptotic Capacity at High SNR}\label{sec:asym_cap}

We denote the capacity with peak power constraint $\snr$ by
$C(\snr)$. For any $n\in\mathbb{N}$ and $\snr>0$, let
\begin{eqnarray*}
\mathbb{D}_{n}(\snr)=\left\{x^n\in \mathbb{C}^n: \max\limits_{1\leq t\leq n} |x_t|^2\; \leq \; \snr\right\}.
\end{eqnarray*}
Let $\mathcal{P}_n(\snr)$ be the set of probability distributions
on $\mathbb{D}_{n}(\snr)$. Since the channel is block-wise
stationary and ergodic, a coding theorem exists and we have
\begin{eqnarray*}
C(\snr)=\lim\limits_{n\rightarrow\infty}\sup\limits_{P_{\xb^n}\in\mathcal{P}_n(\snr)}\frac{1}{n}I(\xb^n;
\yb^n).
\end{eqnarray*}
\subsection{Lower Bound and Upper Bound}
To derive a lower bound on $C(\snr)$ for the channel model given
in (\ref{channelmodel}), we adopt the interleaved
decision-oriented training scheme proposed in \cite{ET06} with
some modifications. This scheme can also be viewed as a way
of interpreting the computations in \cite[Sec. IV.E]{LM03},
\cite[Sec. V]{L05}.

Let $p(\xb)$ be a circularly symmetric distribution with
$\|\xb\|^2\in[x^2_{\min},\snr]$. Construct the codebook
$\mathcal{C}=\mathcal{C}_1\times\mathcal{C}_2\times\cdots\times\mathcal{C}_K$
with $K$ sub-codebooks
$\mathcal{C}_1,\mathcal{C}_2,\cdots,\mathcal{C}_K$, where codebook
$\mathcal{C}_i$ ($i=1,\cdots,K$) contains $2^{nR_i}$ codewords of
length $n$ generated independently symbol by symbol using
distribution $p(\xb)$. We assume that $K$ is a multiple of the
block length $T$, i.e., $K=rT$ for some positive integer $r$.

Now we multiplex (or interleave) these $K$ codebooks.
Specifically, codebook $\mathcal{C}_i$ ($i=1,\cdots,K$) is used at
time instants $i, i+K, i+2K,\cdots,i+(n-1)K$. For codebook
$\mathcal{C}_i$, its codeword can be successfully decoded if
\begin{eqnarray*}
R_1\leq
\frac{1}{n}I\left(\{\xb_{1+jK}\}_{j=0}^{n-1};\{\yb_{1+jK}\}_{j=0}^{n-1}\right)
\end{eqnarray*}
for sufficiently large $n$. Furthermore, using the facts that
$\{\xb_{1+jK}\}_{j=0}^{n-1}$ are i.i.d. and that the channel is
stationary over time instants $1, 1+K, 1+2K,\cdots,1+(n-1)K$, we
get
\begin{eqnarray}
\frac{1}{n}I\left(\{\xb_{1+jK}\}_{j=0}^{n-1};\{\yb_{1+jK}\}_{j=0}^{n-1}\right)&=&\frac{1}{n}\sum\limits_{j=0}^{n-1}I\left(\xb_{1+jK};\{\yb_{1+iK}\}_{i=0}^{n-1}|\{\xb_{1+iK}\}_{i=j+1}^{n-1}\right)\nonumber\\
&=&\frac{1}{n}\sum\limits_{j=0}^{n-1}I\left(\xb_{1+jK};\{\yb_{1+iK}\}_{i=0}^{n-1},\{\xb_{1+iK}\}_{i=j+1}^{n-1}\right)\nonumber\\
&\geq&\frac{1}{n}\sum\limits_{j=0}^{n-1}I(\xb_{1+jK};\yb_{1+jK})\nonumber\\
&=&I(\xb_1;\yb_1) \label{memoryhelp}
\end{eqnarray}
This is to be expected since a channel with memory has a
higher reliable communication rate than the memoryless channel
with the same marginal transition probability. Thus, reliable communication at rate
$R_1=I(\xb_1;\yb_1)$ is possible for sub-codebook
$\mathcal{C}_1$. After
$\{\xb_{1+jK}\}_{j=0}^{n-1}$ is successfully decoded, the receiver
can use these values as well as $\{\yb_{1+jK}\}_{j=0}^{n-1}$ to estimate
$\{\hb_{2+jK}\}_{j=0}^{n-1}$. Specifically,
$(\xb_{1+jK},\yb_{1+jK})$ is used to estimate $\hb_{2+jK}$,
$j=0,1,\cdots,n-1$. Since $\|\xb\|^2\in[x^2_{\min},\snr]$, it is
easy to verify the following Markov chain condition
\[
\hb_{1+jK}+\frac{1}{x_{\min}}\zb_{1+jK}\rightarrow(\xb_{1+jK},\yb_{1+jK})\rightarrow\hb_{2+jK}\: .
\]
To facilitate the calculation, we assume that
$\hb_{1+jK}+\frac{1}{x_{\min}}\zb_{1+jK}$ is used to estimate
$\hb_{2+jK}$ by forming the MMSE estimate
$\mathbb{E}\left(\hb_{2+jK}\left|\hb_{1+jK}+\frac{1}{x_{\min}}\zb_{1+jK}\right.\right)$,
$j=0,1,\cdots,n-1$. The receiver decodes the codeword in codebook
$\mathcal{C}_2$ using
$\left\{\mathbb{E}\left(\hb_{2+jK}\left|\hb_{1+jK}+\frac{1}{x_{\min}}\zb_{1+jK}\right.\right)\right\}_{j=0}^{n-1}$
as side information. Successful decoding is possible if
\begin{eqnarray*}
R_2\leq
\frac{1}{n}\sum\limits_{j=0}^{n-1}I\left(\{\xb_{2+jK}\}_{j=0}^{n-1};\{\yb_{2+jK}\}_{j=0}^{n-1},
\left\{\mathbb{E}\left(\hb_{2+jK}\left|\hb_{1+jK}+\frac{1}{x_{\min}}\zb_{1+jK}\right.\right)\right\}_{j=0}^{n-1}\right).
\end{eqnarray*}
Similar to (\ref{memoryhelp}), we can use the lower bound
\begin{eqnarray*}
&&\frac{1}{n}\sum\limits_{j=0}^{n-1}I\left(\{\xb_{2+jK}\}_{j=0}^{n-1};\{\yb_{2+jK}\}_{j=0}^{n-1},
\left\{\mathbb{E}\left(\hb_{2+jK}\left|\hb_{1+jK}+\frac{1}{x_{\min}}\zb_{1+jK}\right.\right)\right\}_{j=0}^{n-1}\right)\\
&\geq&
I\left(\xb_2;\yb_2,\mathbb{E}\left(\hb_{2}\left|\hb_{1}+\frac{1}{x_{\min}}\zb_{1}\right.\right)\right)\\
&=&I\left(\xb_2;\yb_2\left|\mathbb{E}\left(\hb_{2}\left|\hb_{1}+\frac{1}{x_{\min}}\zb_{1}\right.\right)\right.\right)
\end{eqnarray*}
to show that reliable communication at rate
$R_2=I\left(\xb_2;\yb_2\left|\mathbb{E}\left(\hb_{2}\left|\hb_{1}+\frac{1}{x_{\min}}\zb_{1}\right.\right)\right.\right)$
is possible for sub-codebook $\mathcal{C}_2$. By applying this
procedure successively, we can conclude that for codebook
$\mathcal{C}_i$, reliable communication is possible at rate
\begin{eqnarray*}
R_i=I\left(\xb_i;\yb_i\left|\mathbb{E}\left(\hb_i\left|\left\{\hb_j+\frac{1}{x_{\min}}\zb_j\right\}_{j=1}^{i-1}\right.\right)\right.\right),\quad
i=1,2,\cdots,K.
\end{eqnarray*}
Thus, using this interleaved decision-oriented training scheme, we can have reliable communication at overall rate of
\begin{eqnarray*}
R=\frac{1}{K}\sum\limits_{i=1}^KR_i=\frac{1}{K}\sum\limits_{i=1}^KI\left(\xb_i;\yb_i\left|\mathbb{E}\left(\hb_i\left|\left\{\hb_j+\frac{1}{x_{\min}}\zb_j\right\}_{j=1}^{i-1}\right.\right)\right.\right).
\end{eqnarray*}

We show in Appendix \ref{app:monotonicity} that
$\left\{I\left(\xb_{i+jT};\yb_{i+jT}\left|\mathbb{E}\left(\hb_{i+jT}\left|\left\{\hb_k+\frac{1}{x_{\min}}\zb_k\right\}_{k=1}^{i+jT-1}\right.\right)\right.\right)\right\}_j$
is a monotone increasing sequence with
\begin{eqnarray*}
\lim\limits_{j\rightarrow\infty}I\left(\xb_{i+jT};\yb_{i+jT}\left|\mathbb{E}\left(\hb_{i+jT}\left|\left\{\hb_k+\frac{1}{x_{\min}}\zb_k\right\}_{k=1}^{i+jT-1}\right.\right)\right.\right)=I\left(\xb_{i};\yb_{i}\left|\mathbb{E}\left(\hb_{i}\left|\left\{\hb_k+\frac{1}{x_{\min}}\zb_k\right\}_{k=-\infty}^{i-1}\right.\right)\right.\right).
\end{eqnarray*}
Now we let $K$ go to infinity  (i.e., we let $r\rightarrow\infty$ since $T$
is fixed), and we obtain
\begin{eqnarray*}
\lim\limits_{K\rightarrow\infty} R&=&\lim\limits_{K\rightarrow\infty}\frac{1}{K}\sum\limits_{i=1}^KI\left(\xb_i;\yb_i\left|\mathbb{E}\left(\hb_i\left|\left\{\hb_j+\frac{1}{x_{\min}}\zb_j\right\}_{j=1}^{i-1}\right.\right)\right.\right)\\
&=&\lim\limits_{K\rightarrow\infty}\frac{1}{K}\sum\limits_{i=1}^T\sum\limits_{j=0}^{r-1}I\left(\xb_{i+jT};\yb_{i+jT}\left|\mathbb{E}\left(\hb_{i+jT}\left|\left\{\hb_k+\frac{1}{x_{\min}}\zb_k\right\}_{k=1}^{i+jT-1}\right.\right)\right.\right)\\
&=&\frac{1}{T}\sum\limits_{i=1}^T\left[\lim\limits_{r\rightarrow\infty}\frac{1}{r}\sum\limits_{j=0}^{r-1}I\left(\xb_{i+jT};\yb_{i+jT}\left|\mathbb{E}\left(\hb_{i+jT}\left|\left\{\hb_k+\frac{1}{x_{\min}}\zb_k\right\}_{k=1}^{i+jT-1}\right.\right)\right.\right)\right] \\
&=&\frac{1}{T}\sum\limits_{i=1}^TI\left(\xb_{i};\yb_{i}\left|\mathbb{E}\left(\hb_{i}\left|\left\{\hb_k+\frac{1}{x_{\min}}\zb_k\right\}_{k=-\infty}^{i-1}\right.\right)\right.\right)\\
&=&\frac{1}{T}\sum\limits_{i=1}^TI\left(\xb_{i};\hb_{i}\xb_i+\zb_i\left|\mathbb{E}\left(\hb_{i}\left|\left\{\hb_k+\frac{1}{x_{\min}}\zb_k\right\}_{k=-\infty}^{i-1}\right.\right)\right.\right).
\end{eqnarray*}
This yields the single-letter lower bound
\begin{eqnarray}
C(\snr)\geq\frac{1}{T}\sum\limits_{i=1}^TI\left(\xb_{i};\hb_i\xb_i+\zb_i\left|\mathbb{E}\left(\hb_{i}\left|\left\{\hb_k+\frac{1}{x_{\min}}\zb_k\right\}_{k=-\infty}^{i-1}\right.\right)\right.\right) \label{lowerbound}
\end{eqnarray}
where $\xb_1,\xb_2,\cdots,\xb_T$ all have the same distribution
$p(\xb)$, which is to be optimized later.

\begin{remark}
Although channel estimation and communication are
intertwined in this interleaved decision-oriented training scheme,
the effect of channel memory is isolated from channel coding
through interleaving. This is because when $K$ is large enough,
$\hb_{i},\hb_{i+K},\hb_{i+2K},\cdots,\hb_{i+(n-1)K}$ are roughly
independent. Thus the codeword in codebook $\mathcal{C}_i$, which is
transmitted over time instants $i,i+K,i+2K,\cdots,i+(n-1)K$,
essentially experiences a memoryless channel. This also suggests
that as $K$ goes to infinity, the single-letter lower bound
(\ref{lowerbound}) provides a correct estimate of the rate
supported by this interleaved decision-oriented training scheme.
We can see that the channel
memory manifests itself in the lower bound (\ref{lowerbound})  only through
$\mathbb{E}\left(\hb_{i}\left|\left\{\hb_k+\frac{1}{x_{\min}}\zb_k\right\}_{k=-\infty}^{i-1}\right.\right)$.
Furthermore, in (\ref{lowerbound}), we can write $\hb_i$ as the
sum of two independent random variables: the coherent fading
component
$\mathbb{E}\left(\hb_{i}\left|\left\{\hb_k+\frac{1}{x_{\min}}\zb_k\right\}_{k=-\infty}^{i-1}\right.\right)$
which is known the the receiver, and the non-coherent fading
component
$\hb_i-\mathbb{E}\left(\hb_{i}\left|\left\{\hb_k+\frac{1}{x_{\min}}\zb_k\right\}_{k=-\infty}^{i-1}\right.\right)$
which is unknown. Isolating the effect of channel memory
facilitates the channel code design: we only need to design
channel codes for memoryless fading channels with different
coherent and non-coherent components, instead of designing
different codes for channels with different memory structures.
\end{remark}

To derive a single-letter upper bound on $C(\snr)$, we follow the
approach in \cite{L05}. The capacity $C(\snr)$ is given by
\begin{eqnarray*}
C(\snr)=\lim\limits_{n\rightarrow\infty}\sup\limits_{P_{\xb^n}\in\mathcal{P}_n(\snr)}\frac{1}{n}I(\xb^n;
\yb^n).
\end{eqnarray*}
By the chain rule,
\begin{eqnarray*}
I(\xb^n;\yb^n)=\sum\limits_{k=1}^nI(\xb^n;\yb_k|\yb^{k-1}).
\end{eqnarray*}
We can upper-bound $I(\xb^n;\yb_k|\yb^{k-1})$ as
\begin{eqnarray}
I(\xb^n;\yb_k|\yb^{k-1})&=&I(\xb^n,\yb^{k-1};\yb_k)-I(\yb_k;\yb^{k-1})\nonumber\\
&\leq&I(\xb^n,\yb^{k-1};\yb_k)\nonumber\\
&=&I(\xb^k,\yb^{k-1};\yb_k)\nonumber\\
&\leq&I\left(\xb_k,\hb_{k-1}+\frac{1}{\sqrt{\snr}}\zb_{k-1},\cdots,\hb_1+\frac{1}{\sqrt{\snr}}\zb_{1};\yb_k\right)\label{lateruse}\\
&\leq&I\left(\xb_k,\left\{\hb_t+\frac{1}{\sqrt{\snr}}\zb_{t}\right\}_{t=-\infty}^{k-1};\yb_k\right).\label{lateruse2}
\end{eqnarray}
Since
$\left(\xb_k,\mathbb{E}\left(\hb_k\left|\left\{\hb_t+\frac{1}{\sqrt{\snr}}\zb_{t}\right\}_{t=-\infty}^{k-1}\right.\right)\right)$
is a sufficient statistic for estimating $\yb_k$ from
$\left(\xb_k,\left\{\hb_t+\frac{1}{\sqrt{\snr}}\zb_{t}\right\}_{t=-\infty}^{k-1}\right)$,
it follows that
\begin{eqnarray*}
I\left(\xb_k,\left\{\hb_t+\frac{1}{\sqrt{\snr}}\zb_{t}\right\}_{t=-\infty}^{k-1};\yb_k\right)=I\left(\xb_k,\mathbb{E}\left(\hb_k\left|\left\{\hb_t+\frac{1}{\sqrt{\snr}}\zb_{t}\right\}_{t=-\infty}^{k-1}\right.\right);\yb_k\right).
\end{eqnarray*}
Note that by the block stationarity of the fading process,
\begin{eqnarray*}
I\left(\xb_k,\mathbb{E}\left(\hb_k\left|\left\{\hb_t+\frac{1}{\sqrt{\snr}}\zb_{t}\right\}_{t=-\infty}^{k-1}\right.\right);\yb_k\right)
\end{eqnarray*}
depends on $k$ only through ($k$ mod $T$). Therefore, we obtain the
single-letter upper bound
\begin{eqnarray}
C(\snr)&\leq&\frac{1}{T}\sum\limits_{k=1}^T\sup\limits_{P_{\xb_k}\in\mathcal{P}_1(\snr)}I\left(\xb_k,\mathbb{E}\left(\hb_k\left|\left\{\hb_t+\frac{1}{\sqrt{\snr}}\zb_{t}\right\}_{t=-\infty}^{k-1}\right.\right);\yb_k\right).\label{upperbound}
\end{eqnarray}

\subsection{Asymptotic Analysis}

Now we proceed to show that the lower bound (\ref{lowerbound}) and
upper bound (\ref{upperbound}) together characterize the
asymptotic behavior of $C(\snr)$ in the high SNR regime.

\begin{lemma}\label{lemma:rank}
For every $\xi\in [\xi_0,\xi_1]$, let $A(\xi)$ be an $M\times M$
symmetric positive semidefinite matrix. We have
\begin{eqnarray*}
\lim\limits_{\epsilon\rightarrow 0
}\frac{\int_{\xi_0}^{\xi_1}\log\det\left[A(\xi)+\epsilon
I_M\right]\mbox{d}\xi}{\log\epsilon}=\sum\limits_{i=0}^{M}
(M-i)\mu\left(\mbox{rank}\left(A(\xi)\right)=i\right).
\end{eqnarray*}
where $\mu(\mbox{rank}(A(\xi))=i)$ is the Lebesgue measure of the
set $\{\xi:\mbox{rank}(A(\xi))=i\}$.

For the special case where $A(\xi)=A$ for all
$\xi\in[\xi_0,\xi_1]$ and $\xi_1-\xi_0=1$, we get
\begin{eqnarray*}
\lim\limits_{\epsilon\rightarrow 0 }\frac{\log\det\left[A+\epsilon
I_M\right]}{\log\epsilon}=M-\mbox{rank}(A).
\end{eqnarray*}
\end{lemma}
\begin{proof}
See Appendix \ref{app:rank}.
\end{proof}

\begin{lemma}[\cite{L05}, Sec. V]\label{lemma:computation}
If $\xb$ is uniformly distributed over the set
$\left\{z\in\mathbb{C}:\frac{\sqrt{\snr}}{2}\leq|z|\leq\sqrt{\snr}\right\}$,
$\widehat{\hb}\sim\mathcal{CN}\left(0,\mathbb{E}\left|\widehat{\hb}\right|^2\right)$,
$\widetilde{\hb}\sim\mathcal{CN}\left(0,\mathbb{E}\left|\widetilde{\hb}\right|^2\right)$,
$\zb\sim\mathcal{CN}(0,1)$, and
$\xb,\widehat{\hb},\widetilde{\hb},\widehat{\zb}$ are all
independent, then
\begin{eqnarray}
I\left(\xb;\left.\left(\widehat{\hb}+\widetilde{\hb}\right)\xb+\zb\right|\widehat{\hb}\right)\geq-\log\left(\mathbb{E}\left|\widetilde{\hb}\right|^2+\frac{8}{5\snr}\right)+\log\left(1-\mathbb{E}\left|\widetilde{\hb}\right|^2\right)-\gamma-\log\frac{5e}{6}\label{comput}
\end{eqnarray}
where $\gamma$ is the Euler constant.
\end{lemma}

\begin{theorem} \label{th:asym_cap}
For the block-stationary Gaussian fading channel model given in
(\ref{channelmodel}),
\begin{eqnarray}
\lim\limits_{\snr\rightarrow\infty}\frac{
C(\snr)}{\log\snr}=\lim\limits_{\snr\rightarrow\infty}\frac{-\log\det\left[\Sigma(\snr)\right]}{T\log\snr}=\frac{1}{2\pi
T}\sum\limits_{i=0}^{T}
(T-i)\mu\left(\mbox{rank}(S(e^{j\omega}))=i\right). \label{prelog}
\end{eqnarray}
\end{theorem}
\begin{remark}
The second equality in (\ref{prelog}) follows from
(\ref{identity2}) and Lemma \ref{lemma:rank}.
\end{remark}

\begin{proof}
Below we provide an intuitive explanation of this theorem based on the
lower bound (\ref{lowerbound}). The details of the proof are left
to Appendix \ref{app:asym_cap}.

In the lower bound (\ref{lowerbound}), let $\xb_i$ be uniformly
distributed over the set
$\left\{z\in\mathbb{C}:\frac{\sqrt{\snr}}{2}\leq|z|\leq\sqrt{\snr}\right\}$, and
write $\hb_i$ as $\hb_i=\widehat{\hb}_i+\widetilde{\hb}_i$ where
$\widehat{\hb}_i=\mathbb{E}\left(\hb_i\left|\left\{\hb_k+\frac{2}{\sqrt{\snr}}\zb_k\right\}_{k=-\infty}^{i-1}\right.\right)$.
Suppose
$\mathbb{E}\left|\widetilde{\hb}_i\right|^2\approx\snr^{-r_i}$,
$i=1,2,\cdots,T$. We can then write $\yb_i$ as
$\yb_i=\widehat{\hb}_i\xb_i+\wb_i$ where
$\wb_i=\widetilde{\hb}_i\xb_i+\zb_i$ with
$\mathbb{E}\left|\wb_i\right|^2\approx\snr^{1-r_i}$. By viewing
$\yb_i$ as the output of a coherent fading channel with the fading
$\widehat{\hb}_i$ known at the receiver and noise $\wb_i$, we get
\begin{eqnarray*}
I\left(\xb_i;\hb_i\xb_i+\zb_i\left|\mathbb{E}\left(\hb_i\left|\left\{\hb_k+\frac{2}{\sqrt{\snr}}\zb_k\right\}_{k=-\infty}^{i-1}\right.\right)\right.\right)&=&I\left(\xb_i;\widehat{\hb}_i\xb_i+\wb_i\left|\widehat{\hb}_i\right.\right)\\
&\approx&\log\frac{\snr}{\snr^{1-r_i}}\\
&=&r_i\log\snr.
\end{eqnarray*}
Thus the lower bound (\ref{lowerbound}) can be approximated by
\begin{eqnarray*}
\frac{1}{T}\sum\limits_{i=1}^TI\left(\xb_i;\hb_i\xb_i+\zb_i\left|\mathbb{E}\left(\hb_i\left|\left\{\hb_k+\frac{2}{\sqrt{\snr}}\zb_k\right\}_{k=-\infty}^{i-1}\right.\right)\right.\right)\approx\frac{1}{T}\sum\limits_{i=1}^Tr_i\log\snr.
\end{eqnarray*}
We then complete the proof by showing that $\sum_{i=1}^Tr_i$ is
related to the matrix spectral density function $S(e^{j\omega})$
by the equation
\begin{eqnarray*}
\sum\limits_{i=1}^Tr_i=\frac{1}{2\pi}\sum\limits_{i=0}^{T}
(T-i)\mu\left(\mbox{rank}(S(e^{j\omega}))=i\right).
\end{eqnarray*}
\end{proof}

Theorem~\ref{th:asym_cap} generalizes many previous results on the
noncoherent capacity for Gaussian channels in the high SNR regime
as we illustrate in the following subsection.

\subsection{Previous Results as Special Cases of Theorem~\ref{th:asym_cap}}
\begin{example} \textsl{Constant Fading within Block}

For the special case where the fading remains constant within a
block, i.e., $\hb_{kT+1}=\hb_{kT+2}=\cdots=\hb_{kT+T}$, for all
$k\in\mathbb{Z}$, all the entries of $R(i)$ for any fixed $i$ are
identical. This implies that, for any fixed $\omega$, all the
entries of $S(e^{j\omega})$ are identical, which we shall denote
by $s(e^{j\omega})$. It is easy to see that $s(e^{j\omega})$ is
essentially the spectral density function of $\{\hb_{kT}\}_k$. The
rank of $S(e^{j\omega})$ is 1 if $s(e^{j\omega})>0$, and is 0 if
$s(e^{j\omega})=0$. We therefore have
\begin{eqnarray*}
\lim\limits_{\snr\rightarrow\infty}\frac{C(\snr)}{\log\snr}&=&\frac{1}{2\pi
T}\sum\limits_{i=0}^T(T-i)\mu(\mbox{rank}(S(e^{j\omega}))=i)\\
&=&\frac{1}{2\pi
T}\sum\limits_{i=0}^1(T-i)\mu(\mbox{rank}(S(e^{j\omega}))=i)\\
&=&1-\frac{\mu(s(e^{j\omega})>0)}{2\pi T}.
\end{eqnarray*}
When $T=1$, we recover the result in \cite{L05} that
\begin{eqnarray}
\lim\limits_{\snr\rightarrow\infty}\frac{C(\snr)}{\log\snr}=\frac{\mu(s(e^{j\omega})=0)}{2\pi}
\label{symbolprelog}
\end{eqnarray}
which illustrates the effect of large timescale correlation of the
fading process on the pre-log term of the channel capacity in the
high SNR regime. When the fading is independent from block to
block, we have $\mu(s(e^{j\omega})>0)=2\pi$, and thus recover the
result in \cite{MH99,ZT02} that
\begin{eqnarray*}
\lim\limits_{\snr\rightarrow\infty}\frac{C(\snr)}{\log\snr}=\frac{T-1}{T}
\end{eqnarray*}
which illustrates the effect of short timescale correlation of the
fading process on the pre-log term of the capacity at high SNR.
\end{example}

\begin{example} \textsl{Time-Selectivity within Block}

In this example, we recover the main result in \cite{LV04}
concerning the case where rank deficiency is caused purely by the
correlation within a block.

If $\mbox{rank}(\Sigma(\infty))=\mbox{rank}(R(0))$, then\footnote{The condition
$\mbox{rank}(\Sigma(\infty))=\mbox{rank}(R(0))$ is satisfied, for instance, when
the fading process is independent from block to block.}
\begin{eqnarray}
\lim\limits_{\snr\rightarrow\infty}\frac{C(\snr)}{\log\snr}=\frac{T-\mbox{rank}(R(0))}{T}.\label{Liangprelog}
\end{eqnarray}

To prove (\ref{Liangprelog}), we first note that
\begin{eqnarray*}
\Sigma(\infty)+\frac{1}{\snr}I_T&=&\mbox{var}\left(\left.\left(\hb_1,\hb_2,\cdots,\hb_T\right)^\top\right|\left\{\hb_t\right\}_{k=-\infty}^0\right)+\frac{1}{\snr}I_T\\
&=&\mbox{var}\left(\left.\left(\hb_1+\frac{1}{\sqrt{\snr}}\zb_1,\cdots,\hb_T+\frac{1}{\sqrt{\snr}}\zb_T\right)^\top\right|\{\hb_k\}_{k=-\infty}^0\right)\\
&\preccurlyeq&
\mbox{var}\left(\left.\left(\hb_1+\frac{1}{\sqrt{\snr}}\zb_1,\cdots,\hb_T+\frac{1}{\sqrt{\snr}}\zb_T\right)^\top\right|\left\{\hb_k+\frac{1}{\sqrt{\snr}}\zb_k\right\}_{k=-\infty}^0\right)\\
&=&\Sigma(\snr)\\
&\preccurlyeq&\mbox{var}\left(\left(\hb_1+\frac{1}{\sqrt{\snr}}\zb_1,\cdots,\hb_T+\frac{1}{\sqrt{\snr}}\zb_T\right)^\top\right)\\
&=&R(0)+\frac{1}{\snr}I_T.
\end{eqnarray*}
We therefore have the bound
\begin{eqnarray*}
\lim\limits_{\snr\rightarrow\infty}\frac{\log\det\left[\Sigma(\infty)+\frac{1}{\snr}I_T\right]}{\log\snr}\leq\lim\limits_{\snr\rightarrow\infty}\frac{\log\det\Sigma(\snr)}{\log\snr}\leq\lim\limits_{\snr\rightarrow\infty}\frac{\log\det\left[R(0)+\frac{1}{\snr}I_T\right]}{\log\snr}.
\end{eqnarray*}
By Lemma \ref{lemma:rank},
\begin{eqnarray*}
\lim\limits_{\snr\rightarrow\infty}\frac{\log\det\left[\Sigma(\infty)+\frac{1}{\snr}I_T\right]}{\log\snr}&=&\lim\limits_{\snr\rightarrow\infty}\frac{\log\det\left[R(0)+\frac{1}{\snr}I_T\right]}{\log\snr}\\
&=&-T+\mbox{rank}(R(0)
\end{eqnarray*}
which implies that
\begin{eqnarray*}
\lim\limits_{\snr\rightarrow\infty}\frac{\log\det\Sigma(\snr)}{\log\snr}=-T+\mbox{rank}(R(0).
\end{eqnarray*}
Therefore, by Theorem \ref{th:asym_cap},
\begin{eqnarray*}
\lim\limits_{\snr\rightarrow\infty}\frac{C(\snr)}{\log\snr}&=&\lim\limits_{\snr\rightarrow\infty}\frac{-\log\det\Sigma(\snr)}{T\log\snr}\\
&=&\frac{T-\mbox{rank}(R(0))}{T}.
\end{eqnarray*}

It is worth noting that in this case the pre-log term of the capacity
can be achieved by a scheme simpler than the aforementioned
interleaved decision-oriented training scheme. Suppose the rank of
$R(0)$ is $Q$, so  that $R(0)$ has $Q\times Q$ positive definite
principal submatrix. Without loss of generality, suppose this
submatrix is the covariance matrix of
$(\hb_1,\hb_2,\cdots,\hb_Q)^T$. Then $\hb_{kT+i}$ can be
represented as a linear combination of
$\hb_{kT+1},\hb_{kT_2},\cdots,\hb_{kT+Q}$ for any $k\in\mathbb{Z}$
and $i\in\{Q+1,Q+2,\cdots,T\}$. The simpler scheme is described as
follows:

The transmitter sends deterministic training symbols with maximum
power at time instants $kT+1,kT+2,\cdots,kT+Q$, i.e.,
$\xb_{kT+1}=\xb_{kT+2}=\cdots=\xb_{kT+Q}=\sqrt{\snr}$, where
$k=0,1,2,\cdots$. The receiver can form the MMSE estimates
$\mathbb{E}\left(\hb_{kT+i}\left|\left\{\hb_{kT+j}+\frac{1}{\sqrt{\snr}}\zb_{kT+j}\right\}_{j=1}^Q\right.\right)$,
for $i=Q+1,Q+2,\cdots,T$ and $k=0,1,2,\cdots$. Clearly, we have
\begin{eqnarray*}
\mbox{var}\left(\hb_{kT+i}\left|\left\{\hb_{kT+j}+\frac{1}{\sqrt{\snr}}\zb_{kT+j}\right\}_{j=1}^Q\right.\right)=O\left(\frac{1}{\snr}\right).
\end{eqnarray*}
With the side information
$\left\{\mathbb{E}\left(\hb_{kT+i}\left|\left\{\hb_{kT+j}+\frac{1}{\sqrt{\snr}}\zb_{kT+j}\right\}_{j=1}^Q\right.\right)\right\}_{k=0}^\infty$
at the receiver, we can communicate reliably at time instants
$i,T+i,2T+i,\cdots$ with rate at least
$I\left(\xb_i;\yb_i\left|\mathbb{E}\left(\hb_{i}\left|\left\{\hb_{j}+\frac{1}{\sqrt{\snr}}\zb_{j}\right\}_{j=1}^Q\right.\right)\right.\right)$.
Let $\xb_i$ be uniformly distributed over the set
$\{z\in\mathbb{C}:\snr/2\leq\|z\|\leq\snr\}$. By Lemma
\ref{lemma:computation},
\begin{eqnarray*}
I\left(\xb_i;\yb_i\left|\mathbb{E}\left(\hb_{i}\left|\left\{\hb_{j}+\frac{1}{\sqrt{\snr}}\zb_{j}\right\}_{j=1}^Q\right.\right)\right.\right)&\geq&-\log\left[\mbox{var}\left(\hb_{i}\left|\left\{\hb_{j}+\frac{1}{\sqrt{\snr}}\zb_{j}\right\}_{j=1}^Q\right.\right)+\frac{8}{5\snr}\right]\\
&&+\log\left(1-\mbox{var}\left(\hb_{i}\left|\left\{\hb_{j}+\frac{1}{\sqrt{\snr}}\zb_{j}\right\}_{j=1}^Q\right.\right)\right)-\gamma-\log\frac{5e}{6}\\
&=&\log\snr+o(\log\snr).
\end{eqnarray*}
Threfore, the overall rate is lower-bounded by
\begin{eqnarray*}
\frac{1}{T}\sum\limits_{i=Q+1}^TI\left(\xb_i;\yb_i\left|\mathbb{E}\left(\hb_{i}\left|\left\{\hb_{j}+\frac{1}{\sqrt{\snr}}\zb_{j}\right\}_{j=1}^Q\right.\right)\right.\right)&=&\frac{T-Q}{T}\log\snr+o(\log\snr)\\
&=&\frac{T-\mbox{rank}(R(0))}{T}\log\snr+o(\log\snr)
\end{eqnarray*}
and the pre-log term is achieved. This scheme has the following obvious advantages
over the interleaved decision-oriented training scheme: (i) channel
estimation and communication are completely decoupled; and (ii) channel
estimation is done locally since the estimate
$\mathbb{E}\left(\hb_{kT+i}\left|\left\{\hb_{kT+j}+\frac{1}{\sqrt{\snr}}\zb_{kT+j}\right\}_{j=1}^Q\right.\right)$,
$i=Q+1,Q+2,\cdots,T,$ only depends
$\hb_{kT+1},\hb_{kT+2},\cdots,\hb_{kT+Q}$.
\end{example}

\subsection{Regular Block-Stationary Process}

The following theorem generalizes \cite[Corollary 4.42]{LM03} for
regular Gaussian fading processes to the block-stationary case.
\begin{theorem}\label{th:asym_cap2}
If $\det(\Sigma(\infty))>0$, then
\begin{eqnarray}
\lim\limits_{\snr\rightarrow\infty}\left[C(\snr)-\log\log\snr\right]=-1-\gamma-\frac{1}{T}\log\det(\Sigma(\infty))=-1-\gamma-\frac{1}{2\pi
T}\int_{-\pi}^{\pi}\log\det\left[S(e^{j\omega})\right]\mbox{d}\omega.
\label{fadingnum}
\end{eqnarray}
\end{theorem}
\begin{remark}
The second equality in (\ref{fadingnum}) follows from
(\ref{identity1}).
\end{remark}

\begin{proof}
See Appendix \ref{app:asym_cap2}.
\end{proof}

\begin{example}\textsl{Gauss-Markov Process}

Suppose $\{\hb_t\}_{t=-\infty}^\infty$ is a Gauss-Markov process
with $E(\hb_{t+1}\hb^*_{t})=\rho_1$ if $(t \mod T)=0$, and
$=\rho_2$ otherwise. Here $\rho_1,\rho_2$ are complex numbers with
$\max(|\rho_1|,|\rho_2|)<1$. In this case, we have
\begin{eqnarray*}
\det(\Sigma(\infty))=(1-|\rho_1|^2)(1-|\rho_2|^2)^{T-1}.
\end{eqnarray*}
Therefore, by Theorem \ref{th:asym_cap2}
\begin{eqnarray*}
\lim\limits_{\snr\rightarrow\infty}\left[C(\snr)-\log\log\snr\right]=-1-\gamma-\log(1-|\rho_2|^2)-\frac{\log(1-|\rho_1|^2)-\log(1-|\rho_1|^2)}{T}.
\end{eqnarray*}
\end{example}

\section{Symbol-by-Symbol Stationary Fading
Model}\label{sec:discussion}

For simplicity, we assume in this section that the fading
process is symbol-by-symbol stationary, i.e., $T=1$. In this case,
Theorem 1 is specialized to Equation (\ref{symbolprelog}).

\subsection{Best- and Worst-Case Spectral Densities}
We can see that two fading processes with spectral density
functions $s_1(e^{j\omega})$ and $s_2(e^{j\omega})$ can induce the
same pre-log term in the high SNR regime as long as
$\mu(s_1(e^{j\omega})=0)=\mu(s_2(e^{j\omega})=0)$. But in the
non-asymptotic regime, the capacities of these two channels may
behave very differently. So, for a fixed $\mu(s(e^{j\omega})=0)$,
it is natural to ask the question: which spectral density function
$s(e^{j\omega})$ gives the largest (or smallest) channel capacity
at a given $\snr$? This question is difficult to answer since we
do not have a closed-form expression for noncoherent channel
capacity. We therefore turn to the lower bound (\ref{lowerbound})
to formulate a closely-related problem.

When $T=1$, the lower bound
(\ref{lowerbound}) can be reduced to
\begin{eqnarray}
C(\snr)\geq
I\left(\xb_{1};\hb_1\xb_1+\zb_1\left|\mathbb{E}\left(\hb_{1}\left|\left\{\hb_k+\frac{1}{x_{\min}}\zb_1\right\}_{k=-\infty}^{0}\right.\right)\right.\right).\label{lowerboundsym}
\end{eqnarray}
We can see that the lower bound (\ref{lowerboundsym}) depends on
$s(e^{j\omega})$ only through
$\mathbb{E}\left(\hb_{1}\left|\left\{\hb_k+\frac{1}{x_{\min}}\zb_1\right\}_{k=-\infty}^{0}\right.\right)$.
Furthermore, if we fix the input distribution $p(\xb_1)$, then
\begin{eqnarray*}
\mbox{var}\left.\left(\hb_{1}\left|\left\{\hb_k+\frac{1}{x_{\min}}\zb_1\right\}_{k=-\infty}^{0}\right.\right)\right|_{s_1(e^{j\omega})}\leq\mbox{var}\left.\left(\hb_{1}\left|\left\{\hb_k+\frac{1}{x_{\min}}\zb_1\right\}_{k=-\infty}^{0}\right.\right)\right|_{s_2(e^{j\omega})}
\end{eqnarray*}
implies that
\begin{eqnarray*}
&&\left.I\left(\xb_{1};\hb_1\xb_1+\zb_1\left|\mathbb{E}\left(\hb_{1}\left|\left\{\hb_k+\frac{1}{x_{\min}}\zb_1\right\}_{k=-\infty}^{0}\right.\right)\right.\right)\right|_{s_1(e^{j\omega})}\\
&\geq&\left.I\left(\xb_{1};\hb_1\xb_1+\zb_1\left|\mathbb{E}\left(\hb_{1}\left|\left\{\hb_k+\frac{1}{x_{\min}}\zb_1\right\}_{k=-\infty}^{0}\right.\right)\right.\right)\right|_{s_2(e^{j\omega})}.
\end{eqnarray*}
We can therefore ask which $s(e^{j\omega})$ gives the largest (or
smallest)
$\mbox{var}\left.\left(\hb_{1}\left|\left\{\hb_k+\frac{1}{x_{\min}}\zb_1\right\}_{k=-\infty}^{0}\right.\right)\right|_{s(e^{j\omega})}$.
More precisely, since
\begin{eqnarray*}
\mbox{var}\left.\left(\hb_{1}\left|\left\{\hb_k+\frac{1}{x_{\min}}\zb_1\right\}_{k=-\infty}^{0}\right.\right)\right|_{s(e^{j\omega})}&=&\mbox{var}\left.\left(\hb_{1}+\frac{1}{x_{\min}}\zb_1\left|\left\{\hb_k+\frac{1}{x_{\min}}\zb_1\right\}_{k=-\infty}^{0}\right.\right)\right|_{s(e^{j\omega})}-\frac{1}{x^2_{\min}}\\
&=&\exp\left\{\frac{1}{2\pi}\int_{-\pi}^{\pi}\log\left[s(e^{j\omega})+\frac{1}{x^2_{\min}}\right]\mbox{d}\omega\right\}-\frac{1}{x^2_{\min}},
\end{eqnarray*}
we can formulate the problem in the following form:
\begin{eqnarray}
\arg\max_{s(e^{j\omega})} (\mbox{ or } \arg\min_{s(e^{j\omega})})
\int_{-\pi}^{\pi}\log\left[s(e^{j\omega})+\frac{1}{x^2_{\min}}\right]\mbox{d}\omega
\label{optimization}
\end{eqnarray}
subject to
\begin{eqnarray*}
s(e^{j\omega})\geq
0,\quad\frac{1}{2\pi}\int_{-\pi}^{\pi}s(e^{j\omega})\mbox{d}\omega=1,\quad
\mu(s(e^{j\omega})=0)=\alpha.
\end{eqnarray*}
where $\alpha\in[0,2\pi)$. Due to the strict concavity of
$\log(\cdot)$, it is easy to show that the maximizers of the
optimization problem (\ref{optimization}) are the set of spectral
density functions $s(e^{j\omega})$ with the property:
\[
\mu\left(s(e^{j\omega})=\frac{1}{2\pi-\alpha}\right)=2\pi-\alpha,\quad \mu(s(e^{j\omega})=0)=\alpha.
\]
This solution has the following
interpretation. Without constraints on the spectral density
function, the worst fading process is the i.i.d. Gaussian process
whose spectral density function is flat. With the constraint
$\mu(s(e^{j\omega})=0)=\alpha$, the spectral density function
$s(e^{j\omega})$ cannot be completely flat, but the worst fading
process should have a spectral density function  that is as flat as
possible, i.e., the correlation in the time domain
should be the weakest possible. Note that the solution does not
depend on $x_{\min}$. We can use this fact to derive a universal
lower bound on $C(\snr)$ for the class of spectral density
functions $\{s(e^{j\omega}):\mu(s(e^{j\omega})=0)=\alpha\}$, which has further implications for the high SNR asymptotic behavior of $C(\snr)$.
%
%\subsection{A Useful Lower Bound and its Implications}
%
Let
$s_{\max}(e^{j\omega})$ be a maximizer of (\ref{optimization}). We
have
\begin{eqnarray*}
\mbox{var}\left.\left(\hb_{1}\left|\left\{\hb_k+\frac{1}{x_{\min}}\zb_1\right\}_{k=-\infty}^{0}\right.\right)\right|_{s_{\max}(e^{j\omega})}=\left(\frac{1}{2\pi-\alpha}+\frac{1}{x^2_{\min}}\right)^{\frac{2\pi-\alpha}{2\pi}}\left(\frac{1}{x^2_{\min}}\right)^{\frac{\alpha}{2\pi}}-\frac{1}{x^2_{\min}}\triangleq\phi(\alpha,x_{\min}).
\end{eqnarray*}
For any spectral density function $s(e^{j\omega})$ with
$\mu(s(e^{j\omega})=0)=\alpha$, we have
\begin{eqnarray}
\left.C(\snr)\right|_{s(e^{j\omega})}&\geq&
\left.I\left(\xb_{1};\hb_1\xb_1+\zb_1\left|\mathbb{E}\left(\hb_{1}\left|\left\{\hb_k+\frac{1}{x_{\min}}\zb_1\right\}_{k=-\infty}^{0}\right.\right)\right.\right)\right|_{s_{\max}(e^{j\omega})}\nonumber\\
&=&I\left(\xb;\left.\left(\widehat{\hb}+\widetilde{\hb}\right)\xb+\zb\right|\widehat{\hb}\right)\label{universallower}
\end{eqnarray}
where $\xb$ is uniformly distributed over the set
$\left\{z\in\mathbb{C}:x_{\min}\leq|z|\leq\sqrt{\snr}\right\}$,
$\widehat{\hb}\sim\mathcal{CN}\left(0,1-\phi(\alpha,x_{\min})\right)$,
$\widetilde{\hb}\sim\mathcal{CN}\left(0,\phi(\alpha,x_{\min})\right)$,
$\zb\sim\mathcal{CN}(0,1)$, and
$\xb,\widehat{\hb},\widetilde{\hb},\widehat{\zb}$ are all
independent. We can further optimize over $p(\xb)$ to tighten the
lower bound (\ref{universallower}).

The minimizers of (\ref{optimization}) do not exist. Consider the
following set spectral density functions
$\{s_\theta(e^{j\omega})\}_{\theta}$ given by
\begin{eqnarray*}
s_\theta(e^{j\omega})=\left\{\begin{array}{ll} 0 & |\omega|\leq\frac{\alpha}{2}\\
\frac{1}{\theta} & |\omega|\in(\frac{\alpha}{2},\pi-\frac{1}{2\theta}]\\
\frac{2\pi\theta^2-2\pi\theta+\alpha\theta+1}{\theta}
&|\omega|\in(\pi-\frac{1}{2\theta},\pi]
\end{array}\right.&\\
\end{eqnarray*}
where $\theta\geq\theta_0$ with
\begin{eqnarray*}
\theta_0=\left\{\begin{array}{ll} \frac{1}{2\pi-\alpha} & (2\pi-\alpha)^2<8\pi\\
\frac{2\pi-\alpha+\sqrt{(2\pi-\alpha)^2-8\pi}}{4\pi} &
(2\pi-\alpha)^2\geq 8\pi.
\end{array}\right.&\\
\end{eqnarray*}
We can compute
\begin{eqnarray*}
&&\lim\limits_{\theta\rightarrow\infty}\int_{-\pi}^{\pi}\log\left[s_\theta(e^{j\omega})+\frac{1}{x^2_{\min}}\right]\mbox{d}\omega\\
&=&\lim\limits_{\theta\rightarrow\infty}\left[-2\alpha\log x_{\min}+\left(2\pi-\alpha-\frac{1}{\theta}\right)\log\left(\frac{1}{\theta}+\frac{1}{x^2_{\min}}\right)+\frac{1}{\theta}\log\left(\frac{2\pi\theta^2-2\pi\theta+\alpha\theta+1}{\theta}+x^2_{\min}\right)\right]\\
&=&-4\pi\log x_{\min}.
\end{eqnarray*}
Note that
$\int_{-\pi}^{\pi}\log\left[s(e^{j\omega})+\frac{1}{x^2_{\min}}\right]\mbox{d}\omega<-4\pi\log
x_{\min}$. Therefore, as $\theta$ goes to infinity,
$\int_{-\pi}^{\pi}\log\left[s(e^{j\omega})+\frac{1}{x^2_{\min}}\right]\mbox{d}\omega$
approaches the lower bound that is not attainable by any spectral
density function. Intuitively, the fading process associated with
$s_\theta(e^{j\omega})$ becomes more and more deterministic as
$\theta$ gets larger, and it can be verified that
\begin{eqnarray*}
\lim\limits_{\theta\rightarrow\infty}\left.\mbox{var}\left(\hb_1\left|\left\{\hb_k+\frac{1}{x_{\min}}\zb_k\right\}_{k=-\infty}^{0}\right.\right)\right|_{s_\theta(e^{j\omega})}=0.
\end{eqnarray*}
This result has interesting implications for the channel
capacity.

\begin{proposition}\label{pro:nonuniform}
For any $r\geq 0$,
\begin{eqnarray*}
\liminf\limits_{\snr\rightarrow\infty,\theta=\snr^{r}}\frac{\left.C(\snr)\right|_{s_\theta(e^{j\omega})}}{\log\snr}\geq
\frac{\alpha+\min(r,1)(2\pi-\alpha)}{2\pi}.
\end{eqnarray*}
If $r\geq 1$, then
\begin{eqnarray*}
\lim\limits_{\snr\rightarrow\infty,\theta=\snr^{r}}\frac{\left.C(\snr)\right|_{s_\theta(e^{j\omega})}}{\log\snr}=1.
\end{eqnarray*}
\end{proposition}
\begin{proof}
See Appendix \ref{app:nonuniform}
\end{proof}
\begin{remark}
Although by Theorem \ref{th:asym_cap}, for any fixed
$\theta$, the ratio between $C(\snr)|_{s_\theta(e^{j\omega})}$ and
$\log\snr$ converges to $\frac{\alpha}{2\pi}$, Proposition
\ref{pro:nonuniform} says that the convergence is not uniform with
respect to $\theta$. This is intuitively clear because when
$\theta$ is large, we have $s_{\theta}(e^{j\omega})\thickapprox 0$
for $|\omega|\in[0,\pi-\frac{1}{2\theta}]$. Therefore it can be
expected that for a large range of SNR, the channel capacity
$C(\snr)|_{s_{\theta}(e^{j\omega})}$ behaves like
$\left(1-\frac{1}{\pi\theta}\right)\log\snr$, which could be
significantly larger than $\frac{\alpha}{2\pi}\log\snr$. For the
extreme case where $\alpha=0$, by Theorem \ref{th:asym_cap2}, for
any fixed $\theta$, the capacity $C(\snr)|_{s(e^{j\omega})}$ grows
like $\log\log\snr$ at high SNR. But Proposition
\ref{pro:nonuniform} implies that even in this extreme case, the
capacity $C(\snr)|_{s_\theta(e^{j\omega})}$ of some $\theta$ can
grow linearly with $\log\snr$ for a large range of SNR. This
is consistent with the result in \cite{ET06} where is was shown
that for the Gauss-Markov process with $E(\hb_{i+1}\hb^*_i)=\rho$,
the capacity $C(\snr)$ grows like $\log\snr$ for a wide range of
SNR levels if $|\rho|$ is close 1. An intuitive explanation is
that if $|\rho|$ is close 1, the spectrum
\begin{eqnarray*}
s(e^{j\omega})=\frac{1-|\rho|^2}{1-2\mbox{Re}(\rho
e^{- j\omega})+|\rho|^2}
\end{eqnarray*}
is approximately zero for all $\omega$ except those around zero,
and we can expect from equation (\ref{symbolprelog}) that
$C(\snr)$ should grow like $\log\snr$ for a wide range of SNR. But
it should be noted that as opposed to a Gauss-Markov process, a
general Gaussian stationary process cannot be characterized by a
single parameter, and the behavior of $C(\snr)$ can be much more
complicated as shown in the following example.
\end{remark}
\begin{example}\label{ex:twolevel}
Consider the spectral density function
\begin{eqnarray*}
s_\theta(e^{j\omega})=\left\{\begin{array}{ll} \epsilon_1 & |\omega|\leq\pi\alpha_1\\
\epsilon_2 & |\omega|\in(\pi\alpha_1,\pi\alpha_2]\\
\frac{1-\alpha_1\epsilon_1-(\alpha_2-\alpha_1)\epsilon_2}{1-\alpha_2}
&|\omega|\in(\pi\alpha_2,\pi]
\end{array}\right.&\\
\end{eqnarray*}
where $0<\alpha_1<\alpha_2<1$, and $\epsilon_1\ll\epsilon_2\ll 1$
(Note: For two positive numbers $a$ and $b$, $a\ll b$ means
$\frac{b}{a}$ is much greater than 1). We show in Appendix \ref{app:twolevel} that
$\frac{C(\snr)}{\log\snr}$ is approximately equal to $\alpha_2$
for $\snr\leq\frac{1}{\epsilon_2}$, and gradually decreases to
$\alpha_1$ as $\snr$ approaches $\frac{1}{\epsilon_1}$.
%The main
%intuition behind this result is that is that $\frac{C(\snr)}{\log%\snr}$ can be approximated
%by
%$-\frac{\mbox{var}\left(\hb_1\left|\left\{\hb_k+\frac{1}{\sqrt
%{\snr}}\zb_k\right\}_{k=-\infty}^0\right.\right)}{\log\snr}$,
%and the latter one is much easier to analyze.

This example shows that $\frac{C(\snr)}{\log\snr}$ can be highly
SNR dependent. For a regular Gaussian fading process, it may
require unreasonably high (impractical) values of SNR in order for the noncoherent channel
capacity $C(\snr)$ to grow like $\log\log\snr$, and the behavior
of $C(\snr)$ at moderate SNR levels may depend highly on the spectral
density function.

\end{example}

Overall, the above analysis suggests that great caution should be exercised when using the asymptotic results in Theorem~\ref{th:asym_cap} and Theorem~\ref{th:asym_cap2} to approximate
the channel capacity $C(\snr)$ at a finite SNR level.

\subsection{Finite Codeword-length Behavior}
Although the asymptotic capacity results might yield
over-pessimistic approximations such as a $\log\log \snr$ growth with $\snr$, they could also lead to
over-optimistic conclusions. In the capacity analysis, it is
assumed that the codeword is of infinite length. But when the
length of codewords is finite, the situation can be dramatically
different. By Fano's inequality, the communication rate $R$ is
upper-bounded by
\begin{eqnarray*}
R\leq\frac{I(\xb^n;\yb^n)+1}{n(1-P_e)}
\end{eqnarray*}
where $n$ is the codeword length and $P_e$ is the decoding error
probability. Suppose we fix $n$ and $P_e$, and let $\snr$ go to
infinity. For a symbol-by-symbol stationary Gaussian fading
process, even if $\mu(s(e^{j\omega})=0)>0$, the correlation matrix
of the fading process over any finite block length can still be
full-rank. Note that $\frac{1}{n}I(\xb^n;\yb^n)$ is upper-bounded
by the capacity of a block-independent Gaussian fading channel
with the correlation matrix of each block given by
$\mathbb{E}[\hb^n(\hb^n)^\dagger]$. Since
$\mathbb{E}[\hb^n(\hb^n)^\dagger]$ is full rank, it follows from
Theorem \ref{th:asym_cap2} that $\frac{1}{n}I(\xb^n;\yb^n)$, and
hence $R$, grows at most like $\log\log\snr$ as $\snr$ goes to
infinity. Therefore, there is no nontrivial tradeoff between
diversity and multiplexing in the sense of \cite{ZT03}. If we want
$R$ to grow linearly with $\log\snr$ while having the decoding
error probability $P_e$ bounded away from 1, the codeword length
$n$ must scale with $\snr$. It is of interest to determine how
fast the codeword $n$ should scale with $\snr$ in order to
guarantee that the rate  $R$ can grow as $\log\snr$ with the
decoding error probability not approaching 1. More precisely,
letting the rate $R(\snr)$, codeword length $n(\snr)$ and decoding
error probability $P_e(\snr)$ all depend on $\snr$,  we wish to
determine conditions on $n(\snr)$ to guarantee the existence of a
sequence of codebooks (indexed by $\snr$) with rate $R(\snr)$ and
codeword length $n(\snr)$ such that
\begin{eqnarray*}
\liminf\limits_{\snr\rightarrow\infty}\frac{R(\snr)}{\log\snr}\geq
r
\end{eqnarray*}
and
\begin{eqnarray*}
\limsup\limits_{\snr\rightarrow\infty}P_e(\snr)\leq P_e
\end{eqnarray*}
where $r>0$ and $P_e\in(0,1)$.

Now we proceed to derive a necessary condition on the growth rate
of $n(\snr)$. It follows by chain rule that
\begin{eqnarray*}
I(\xb^{n(\snr)};\yb^{n(\snr)})=\sum\limits_{k=1}^{n(\snr)}I\left(\left.\xb^{n(\snr)};\yb_k\right|\yb^{k-1}\right).
\end{eqnarray*}
By (\ref{lateruse}), we can upper-bound
$I\left(\left.\xb^{n(\snr)};\yb_k\right|\yb^{k-1}\right)$ as
\begin{eqnarray*}
&&I\left(\left.\xb^{n(\snr)};\yb_k\right|\yb^{k-1}\right)\\
&\leq&\sup\limits_{P_{\xb_k}\in\mathcal{P}_1(\snr)}I\left(\xb_k,\hb_{k-1}+\frac{1}{\sqrt{\snr}}\zb_{k-1},\cdots,\hb_1+\frac{1}{\sqrt{\snr}}\zb_1;\yb_k\right)\\
&\leq&\sup\limits_{P_{\xb_k}\in\mathcal{P}_1(\snr)}I(\xb_k;\yb_k)+\sup\limits_{P_{\xb_k}\in\mathcal{P}_1(\snr)}I\left(\left.\hb_{k-1}+\frac{1}{\sqrt{\snr}}\zb_{k-1},\cdots,\hb_{1}+\frac{1}{\sqrt{\snr}}\zb_{1};\yb_k\right|\xb_k\right)
\end{eqnarray*}
Since
$\sup_{P_{\xb_k}\in\mathcal{P}_1(\snr)}I(\xb_k;\yb_k)=o(\log\snr)$,
and
\begin{eqnarray*}
&&I\left(\left.\hb_{k-1}+\frac{1}{\sqrt{\snr}}\zb_{k-1},\cdots,\hb_{1}+\frac{1}{\sqrt{\snr}}\zb_{1};\yb_k\right|\xb_k\right)\\
&=&\mathbb{E}\left\{\log\left[\frac{1+|\xb_k|^2}{1+|\xb_k|^2\cdot\mbox{var}\left(\hb_k\left|\left\{\hb_v+\frac{1}{\sqrt{\snr}}\zb_v\right\}_{v=1}^{k-1}\right.\right)}\right]\right\}\\
&\leq&\log\left[\frac{1+\snr}{1+\snr\cdot\mbox{var}\left(\hb_k\left|\left\{\hb_v+\frac{1}{\sqrt{\snr}}\zb_v\right\}_{v=1}^{k-1}\right.\right)}\right]\\%%+I\left(\left.\hb_{0}+\frac{1}{\sqrt{\snr}}\zb_{0},\cdots,\hb_{2-k}+\frac{1}{\sqrt{\snr}}\zb_{2-k};\yb_1\right|\hb_1,\xb_1\right)\\
&=&\log\left(\frac{1+\snr}{\snr}\right)-\log\mbox{var}\left(\hb_k+\frac{1}{\sqrt{\snr}}\zb_k\left|\left\{\hb_v+\frac{1}{\sqrt{\snr}}\zb_v\right\}_{v=1}^{k-1}\right.\right)\\
&\leq&\log\left(\frac{1+\snr}{\snr}\right)-\log\mbox{var}\left(\hb_0+\frac{1}{\sqrt{\snr}}\zb_0\left|\left\{\hb_v+\frac{1}{\sqrt{\snr}}\zb_v\right\}_{v=-n(\snr)}^{-1}\right.\right),
\end{eqnarray*}
it follows by Fano's inequality and the condition
$\limsup_{\snr\rightarrow\infty}P_e(\snr)\leq P_e$ that
\begin{eqnarray*}
\liminf\limits_{\snr\rightarrow\infty}\frac{R(\snr)}{\log\snr}&\leq&\liminf\limits_{\snr\rightarrow\infty}\frac{I(\xb^{n(\snr)};\yb^{n(\snr)})+1}{n(\snr)(1-P_e(\snr))\log\snr}\\
&\leq&\liminf\limits_{\snr\rightarrow\infty}\frac{-\log\mbox{var}\left(\hb_0+\frac{1}{\sqrt{\snr}}\zb_0\left|\left\{\hb_v+\frac{1}{\sqrt{\snr}}\zb_v\right\}_{v=-n(\snr)}^{-1}\right.\right)}{(1-P_e(\snr))\log\snr}\\
&\leq&\liminf\limits_{\snr\rightarrow\infty}\frac{-\log\mbox{var}\left(\hb_0+\frac{1}{\sqrt{\snr}}\zb_0\left|\left\{\hb_v+\frac{1}{\sqrt{\snr}}\zb_v\right\}_{v=-n(\snr)}^{-1}\right.\right)}{(1-P_e)\log\snr}.
\end{eqnarray*}
Therefore, in order for
\begin{eqnarray*}
\liminf\limits_{\snr\rightarrow\infty}\frac{R(\snr)}{\log\snr}\geq
r,
\end{eqnarray*}
we must have
\begin{eqnarray}
\liminf\limits_{\snr\rightarrow\infty}\frac{-\log\mbox{var}\left(\hb_0+\frac{1}{\sqrt{\snr}}\zb_0\left|\left\{\hb_v+\frac{1}{\sqrt{\snr}}\zb_v\right\}_{v=-n(\snr)}^{-1}\right.\right)}{\log\snr}\geq
r(1-P_e). \label{scaling}
\end{eqnarray}
Since
$-\log\mbox{var}\left(\hb_0+\frac{1}{\sqrt{\snr}}\zb_0\left|\left\{\hb_v+\frac{1}{\sqrt{\snr}}\zb_v\right\}_{v=-n}^{-1}\right.\right)$
is a monotone increasing function of $n$, %if we have $n'(\snr)\geq
%n(\snr)$ for all sufficiently large $\snr$, then
%\[
%\liminf\limits_{\snr\rightarrow\infty}\frac{-\log\mbox{var}\left(\hb_0+\frac{1}{\sqrt{\snr}}\zb_0\left|\left\{\hb_v+\frac{1}{\sqrt{\snr}}\zb_v\right\}_{v=-n(\snr)}^{-1}\right.\right)}{\log\snr}>0
%\]
%implies that
%\[
%\liminf\limits_{\snr\rightarrow\infty}\frac{-\log\mbox{var}\left(\hb_0+\frac{1}{\sqrt{\snr}}\zb_0\left|\left\{\hb_v+\frac{1}{\sqrt{\snr}}\zb_v\right\}_{v=-n'(\snr)}^{-1}\right.\right)}{\log\snr}>0.
%\]
it is easy to see (\ref{scaling}) implicitly provides us  with a
lower bound on the scaling rate of $n(\snr)$.

In order to derive an explicit lower bound on the scaling rate of
$n(\snr)$, we need to introduce a concept called {\em transfinite
diameter} \cite{G66}.

\begin{definition}
Let $\mathcal{S}$ be a compact set in the plane. Set
\begin{eqnarray*}
&&V(z_1,\cdots,z_n)=\prod\limits_{j>k}(z_j-z_k)\quad n\geq 2,
\mbox{ }
z_i\in\mathcal{S},\\
&&V_n(\mathcal{S})=\max\limits_{z_1,\cdots,z_n\in\mathcal{S}}|V(z_1,\cdots,z_n)|
\end{eqnarray*}
and
\begin{eqnarray*}
\tau_n(\mathcal{S})=\left[V_n(\mathcal{S})\right]^{\frac{2}{n(n-1)}}.
\end{eqnarray*}
The transfinite diameter of $\mathcal{S}$ is defined by
\begin{eqnarray*}
\tau(\mathcal{S})=\lim\limits_{n\rightarrow\infty}\tau_n(\mathcal{S}).
\end{eqnarray*}
\end{definition}
We need the following facts regarding the transfinite diameter.
\begin{enumerate}
\item[(i)] For two compact sets $\mathcal{S}_1$ and $\mathcal{S}_2$
with $\mathcal{S}_1\subseteq\mathcal{S}_2$, we have
$\tau(\mathcal{S}_1)\leq\tau(\mathcal{S}_2)$.
\item[(ii)] The diameter of the unit circle is 1. More generally, the
diameter of an arc of central angle $\theta$ on the unit circle is
$\sin\left(\frac{\theta}{4}\right)$;
\item[(iii)] The transfinite diameter of any closed proper subset of the unit
circle is less than 1.
\end{enumerate}
A full discussion of the transfinite diameter can be found in
\cite{G66}.

Now return to the original problem. Since
\begin{eqnarray*}
\mbox{var}\left(\hb_0+\frac{1}{\sqrt{\snr}}\zb_0\left|\left\{\hb_v+\frac{1}{\sqrt{\snr}}\zb_v\right\}_{v=-n(\snr)}^{-1}\right.\right)\geq\mbox{var}\left(\hb_0\left|\left\{\hb_v\right\}_{v=-n(\snr)}^{-1}\right.\right),
\end{eqnarray*}
we have
\begin{eqnarray*}
\liminf\limits_{\snr\rightarrow\infty}\frac{-\log\mbox{var}\left(\hb_0+\frac{1}{\sqrt{\snr}}\zb_0\left|\left\{\hb_v+\frac{1}{\sqrt{\snr}}\zb_v\right\}_{v=-n(\snr)}^{-1}\right.\right)}{\log\snr}\leq\liminf\limits_{\snr\rightarrow\infty}\frac{-\log\mbox{var}\left(\hb_0\left|\left\{\hb_v\right\}_{v=-n(\snr)}^{-1}\right.\right)}{\log\snr}.
\end{eqnarray*}
Let $\mathcal{S}=\{e^{j\omega}:s(e^{j\omega})>0\}$. It was shown
in \cite{B85} that if the set $\mathcal{S}$ consists consist of a
finite number of arcs of the unit circle, then
\begin{eqnarray*}
\lim\limits_{n\rightarrow\infty}\left(\mbox{var}\left(\hb_0\left|\left\{\hb_v\right\}_{v=-n}^{-1}\right.\right)\right)^{\frac{1}{n}}=\tau(\mathcal{S}).
\end{eqnarray*}
Under the conditions
\begin{enumerate}
\item[(a)]The set $\mathcal{S}$ consists consist of a
finite number of arcs of the unit circle,
\item[(b)] The set $\mathcal{S}$ is a closed proper subset of the unit
circle,
\end{enumerate}
it can be shown by using Facts (i), (ii) and (iii) that
\begin{eqnarray*}
0<\tau(\mathcal{S})<1.
\end{eqnarray*}
Therefore, under Conditions (a) and (b), we have
\begin{eqnarray}
\liminf\limits_{\snr\rightarrow\infty}\frac{-\log\mbox{var}\left(\hb_0\left|\left\{\hb_v\right\}_{v=-n(\snr)}^{-1}\right.\right)}{\log\snr}&=&\liminf\limits_{\snr\rightarrow\infty}\frac{-\frac{1}{n(\snr)}\log\mbox{var}\left(\hb_0\left|\left\{\hb_v\right\}_{v=-n(\snr)}^{-1}\right.\right)}{\frac{1}{n(\snr)}\log\snr}\nonumber\\
&=&\liminf\limits_{\snr\rightarrow\infty}\frac{-n(\snr)\log\tau(\mathcal{S})}{\log\snr}.\label{scalelower}
\end{eqnarray}
It is clear that in order to guarantee that (\ref{scalelower}) is
greater than or equal to $r(1-P_e)$, we must have
\begin{eqnarray*}
\liminf\limits_{\snr\rightarrow\infty}\frac{n(\snr)}{\log\snr}\geq
-\frac{r(1-P_e)}{\log\tau(\snr)}
\end{eqnarray*}
which is a necessary condition on the scaling rate of $n(\snr)$.

In contrast, we show in Appendix \ref{app:AWGN} and Appendix
\ref{app:Rayleigh} that for the AWGN channel and memoryless
coherent Rayleigh fading channel, it is possible to have the rate
$R(\snr)$ grow linearly with $\log\snr$ with fixed codeword length
$n$ and bounded decoding error probability at high SNR. For these
two cases, to facilitate the calculation, we adopt the average
power constraint. But our main conclusion holds also under the
peak power constraint.

\section{Capacity Per Unit Energy} \label{sec:cap_energy}

In the preceding sections, we focused on the channel capacity in
the high SNR regime. Now we proceed to characterize the behavior
of channel capacity in the low average power regime for the
block-stationary Gaussian fading channel model. To this end, we
shall study the capacity per unit energy  (which is denoted by
$C_p(\snr)$) due to its intrinsic connection with the channel
capacity in this regime. The following theorem provides a
general expression for the capacity per unit energy.

%\begin{definition}
%Rate $R$ is $\epsilon$-achievable per unit energy with peak
%constraint $\snr$ if for every $\delta$, there exists
%$\mathcal{E}_0$ large enough such that if
%$\mathcal{E}\geq\mathcal{E}_0$, then an
%$(n,M,\mathcal{E}/n,\snr,\epsilon)$ code can be found with $\log
%M>\mathcal{E}(R-\delta)$.  Rate $R$ is achievable per unit
%energy if it is $\epsilon$-achievable for all $\epsilon \in (0,1)$.
%Finally, the capacity $C_p(\snr)$ per unit energy is the maximum
%achievable rate per unit energy.
%\end{definition}

\begin{theorem}[\cite{V90,SH05}]
\begin{eqnarray*}
C_p(\snr)=\lim\limits_{n\rightarrow\infty}\sup\limits_{\xb^n\in\mathbb{D}_n(\snr)}\frac{D(p_{\yb^n|\xb^n}\|D(p_{\yb^n|0^n})}{\|\xb^n\|_2^2}.
\end{eqnarray*}
Furthermore, the capacity per unit energy is related to the
capacity by
\begin{eqnarray*}
C_p(\snr)=\sup\limits_{\sfP>0}\frac{C(\sfP,\snr)}{\sfP}=\lim\limits_{\sfP\rightarrow
0}\frac{C(\sfP,\snr)}{\sfP}
\end{eqnarray*}
where $C(\sfP,\snr)$ is the channel capacity with average power
constraint $\sfP$ and peak power constraint $\snr$.
\end{theorem}

The following theorem is an extension of \cite[Proposition
3.1]{SH05} for the symbol-symbol stationary channel model to the
block-stationary model.

\begin{theorem}
For the block-stationary Gaussian fading channel model given in
(\ref{channelmodel}),
\begin{eqnarray*}
C_p(\snr)=1-\frac{1}{2\pi\snr}\min\limits_{\mathcal{M}\subseteq\{1,\cdots,T\}}\Psi(\mathcal{M},\snr)
\end{eqnarray*}
where
\begin{eqnarray*}
\Psi(\mathcal{M},\snr)=\frac{1}{|\mathcal{M}|}\int_{-\pi}^{\pi}\log\det\left[I_{|\mathcal{M}|}+\snr
S_{\mathcal{M}}(e^{j\omega})\right]\mbox{d}\omega,
\end{eqnarray*}
and $S_{\mathcal{M}}(e^{j\omega})$ is an
$|\mathcal{M}|\times|\mathcal{M}|$ principal minor of
$S(e^{j\omega})$ with the indices of columns and rows specified by
$\mathcal{M}$.
\end{theorem}
\begin{proof}
The proof is omitted since it is almost identical to that for the
symbol-by-symbol stationary fading channel \cite{SH05}. The only
difference is that although the capacity per unit energy of the
block-stationary fading channel can be asymptotically achieved by
temporal ON-OFF signaling, we have to determine how to allocate ON
symbols in a block. It can be shown that the optimal allocation
scheme is given by $\mathcal{M}^*$, which is the minimizer of
$\min_{\mathcal{M}\in\{1,\cdots,T\}}\Psi(\mathcal{M})$. Here
$\mathcal{M}^*$ might not be unique.
\end{proof}

$C_p(\snr)$ is a monotonically increasing function of $\snr$. It
is easy to see that $C_p(\snr)$ goes to 1 as
$\snr\rightarrow\infty$, and goes to 0 as $\snr\rightarrow 0$. The
following result provides a more precise characterization of the
convergence behavior.
\begin{corollary}\label{cor:snrasym}
At high $\snr$,
\begin{eqnarray}
C_p(\snr)=1-\min\limits_{\mathcal{M}\subseteq\{1,\cdots,T\}}\frac{\sum\limits_{i=0}^{|\mathcal{M}|}i\mu(\mbox{rank}(S_{\mathcal{M}}(e^{j\omega}))=i)\log\snr}{2\pi|\mathcal{M}|\snr}+o(\frac{\log\snr}{\snr}).\label{highsnrasym}
\end{eqnarray}
At low $\snr$, if
\begin{eqnarray*}
\int_{-\pi}^{\pi}\mbox{tr}\left[S^2(e^{j\omega})\right]\mbox{d}\omega<\infty,
\end{eqnarray*}
then
\begin{eqnarray}
C_p(\snr)=\frac{\snr}{4\pi}\max\limits_{\mathcal{M}\subseteq\{1,\cdots,T\}}\frac{1}{|\mathcal{M}|}\int_{-\pi}^{\pi}\mbox{tr}\left[S_{\mathcal{M}}^2(e^{j\omega})\right]\mbox{d}\omega+o(\snr).\label{lowsnrasym}
\end{eqnarray}
\end{corollary}
\begin{proof}
By Lemma \ref{lemma:rank}, at high $\snr$
\begin{eqnarray*}
\int_{-\pi}^{\pi}\log\det\left[\frac{1}{\snr}I_{|\mathcal{M}|}+S_{\mathcal{M}}(e^{j\omega})\right]\mbox{d}\omega=-\sum\limits_{i=0}^{|\mathcal{M}|}(|\mathcal{M}|-i)\mu(\mbox{rank}(S_{\mathcal{M}}(e^{j\omega}))=i)\log\snr+o(\log\snr).
\end{eqnarray*}
Therefore,
\begin{eqnarray*}
C_p(\snr)&=&1-\frac{1}{2\pi\snr}\min\limits_{\mathcal{M}\subseteq\{1,\cdots,T\}}\Psi(\mathcal{M},\snr)\\
&=&1-\frac{1}{2\pi\snr}\min\limits_{\mathcal{M}\subseteq\{1,\cdots,T\}}\left\{\frac{1}{|\mathcal{M}|}\int_{-\pi}^{\pi}\log\det\left[\frac{1}{\snr}I_{|\mathcal{M}|}+S_{\mathcal{M}}(e^{j\omega})\right]\mbox{d}\omega+2\pi\log\snr\right\}\\
&=&1+\max\limits_{\mathcal{M}\subseteq\{1,\cdots,T\}}\frac{\sum\limits_{i=0}^{|\mathcal{M}|}(|\mathcal{M}|-i)\mu\left(\mbox{rank}(S_{\mathcal{M}}(e^{j\omega}))=i\right)\log\snr}{2\pi|\mathcal{M}|\snr}-\frac{\log\snr}{\snr}+o(\frac{\log\snr}{\snr})\\
&=&1-\min\limits_{\mathcal{M}\subseteq\{1,\cdots,T\}}\frac{\sum\limits_{i=0}^{|\mathcal{M}|}i\mu\left(\mbox{rank}(S_{\mathcal{M}}(e^{j\omega}))=i\right)\log\snr}{2\pi|\mathcal{M}|\snr}+o(\frac{\log\snr}{\snr}).
\end{eqnarray*}
At low $\snr$, using the second-order approximation \cite{BV04},
we obtain
\begin{eqnarray*}
\log\det\left[I_{|\mathcal{M}|}+\snr
S_{\mathcal{M}}(e^{j\omega})\right]=\mbox{tr}[S_{\mathcal{M}}(e^{j\omega})]\snr-\frac{1}{2}\mbox{tr}[S^2_{\mathcal{M}}(e^{j\omega})]\snr^2+o(\snr^2).
\end{eqnarray*}
Therefore,
\begin{eqnarray*}
C_p(\snr)&=&1-\frac{1}{2\pi\snr}\min\limits_{\mathcal{M}\subseteq\{1,\cdots,T\}}\Psi(\mathcal{M},\snr)\\
&=&1-\frac{1}{2\pi}\min\limits_{\mathcal{M}\subseteq\{1,\cdots,T\}}\frac{1}{|\mathcal{M}|}\left\{\int_{-\pi}^{\pi}\mbox{tr}[S_{\mathcal{M}}(e^{j\omega})]\mbox{d}\omega-\frac{1}{2}\int_{-\pi}^{\pi}\mbox{tr}[S^2_{\mathcal{M}}(e^{j\omega})]\snr\mbox{d}\omega\right\}+o(\snr)\\
&=&\frac{\snr}{4\pi}\max\limits_{\mathcal{M}\subseteq\{1,\cdots,T\}}\frac{1}{|\mathcal{M}|}\int_{-\pi}^{\pi}\mbox{tr}[S^2_{\mathcal{M}}(e^{j\omega})]\mbox{d}\omega+o(\snr)
\end{eqnarray*}
where the last equality follows from the fact that
\begin{eqnarray*}
\frac{1}{2\pi|\mathcal{M}|}\int_{-\pi}^{\pi}\mbox{tr}[S_{\mathcal{M}}(e^{j\omega})]\mbox{d}\omega=1.
\end{eqnarray*}
\end{proof}
\begin{remark}
Using the inequality
\begin{eqnarray*}
\log\det\left[I_{|\mathcal{M}|}+\snr
S_{\mathcal{M}}(e^{j\omega})\right]\geq\mbox{tr}[S_{\mathcal{M}}(e^{j\omega})]\snr-\frac{1}{2}\mbox{tr}[S^2_{\mathcal{M}}(e^{j\omega})]\snr^2,
\end{eqnarray*}
we can upper bound $C_p(\snr)$ by
\begin{eqnarray*}
C_p(\snr)\leq\frac{\snr}{4\pi}\max\limits_{\mathcal{M}\subseteq\{1,\cdots,T\}}\frac{1}{|\mathcal{M}|}\int_{-\pi}^{\pi}\mbox{tr}[S^2_{\mathcal{M}}(e^{j\omega})]\mbox{d}\omega.
\end{eqnarray*}
It can be seen from Corollary \ref{cor:snrasym} that this upper
bound is a good approximation of $C_p(\snr)$ in the low $\snr$
regime.
\end{remark}

Now we proceed to compute $C_p(\snr)$ in the following examples.
\begin{example}\label{ex:blockindep}
When the channel changes independently from block to block,
$C_p(\snr)$ is equal to
\begin{eqnarray*}
1-\min\limits_{\mathcal{M}\subseteq\{1,\cdots,T\}}\frac{1}{|\mathcal{M}|\snr}\log\det\left[I_{|\mathcal{M}|}+\snr
R_{\mathcal{M}}(0)\right]
\end{eqnarray*}
where $R_{\mathcal{M}}(0)$ is an
$|\mathcal{M}|\times|\mathcal{M}|$ principal minor of $R(0)$ with
the indices of columns and rows specified by $\mathcal{M}$. If we
further let the fading remain constant within a block, then all
the entries of $R(0)$ are one. It is not difficult to show that
\begin{eqnarray*}
\frac{1}{|\mathcal{M}|\snr}\log\det\left[I_{|\mathcal{M}|}+\snr
R_{\mathcal{M}}(0)\right]=\frac{\log(1+|\mathcal{M}|\snr)}{|\mathcal{M}|\snr}
\end{eqnarray*}
which is minimized when $|\mathcal{M}|=T$, i.e.,
$\mathcal{M}=\{1,2,\cdots,T\}$. So we have
\begin{eqnarray*}
C_p(\snr)=1-\frac{\log(1+T\snr)}{T\snr}
\end{eqnarray*}
as shown in \cite{SH05}.
\end{example}

\begin{example}
Consider the case in which the fading process satisfies the
following conditions
\begin{enumerate}
\item All the
off-diagonal entries of $R(0)$ are equal to $\alpha$, where
$\alpha\in\mathbb{C}$ is a constant;
\item All the entries of $R(i)$ are
equal to $\beta_i$ for any non-zero integer $i$, where
$\beta_i\in\mathbb{C}$ is a constant that depends only on $i$.
\end{enumerate}
We also know that the diagonal entries of $R(0)$ are all one. So
for any fixed $\omega$ ($-\pi\leq\omega\leq\pi$), all the diagonal
entries of $I+\snr S(e^{j\omega})$ are identical, and all the
off-diagonal entries of $I+\snr S(e^{j\omega})$ are identical. It
then follows from Szasz's inequality \cite{HJ85} that for any
$\omega\in[-\pi,\pi]$,
\begin{eqnarray*}
\left\{\det[I_{|\mathcal{M}|}+\snr
S_{\mathcal{M}}(e^{j\omega})\right\}^{\frac{1}{|\mathcal{M}|}}
\end{eqnarray*}
is minimized when $\mathcal{M}=\{1,2,\cdots,T\}$. In this case we therefor have
\begin{eqnarray*}
C_p(\snr)=1-\frac{1}{2\pi
T\snr}\int_{-\pi}^{\pi}\log\det\left(I+\snr
S(e^{j\omega})\right)\mbox{d}\omega.
\end{eqnarray*}

If the fading remains constant within a block, then for any fixed
$\omega$, all the entries of $S(e^{j\omega})$ are identical, which
we shall denote by $s(e^{j\omega})$. It can be shown that
\begin{eqnarray*}
\det\left[I+\snr S(e^{j\omega})\right]=1+T\snr s(e^{j\omega}),
\end{eqnarray*}
which yields
\begin{eqnarray}
C_p(\snr)=1-\frac{1}{2\pi T\snr}\int_{-\pi}^{\pi}\log\left[1+T\snr
s(e^{j\omega})\right]\mbox{d}\omega.\label{TSNR}
\end{eqnarray}
We can see from (\ref{TSNR}) that $C_p(\snr)$ is a monotonically
increasing function of $T$ and $\snr$. Intuitively, as $T$ gets
larger and larger, the receiver can estimate the channel more and
more accurately, and thus the capacity per unit energy of the
non-coherent channel should converge to that of the coherent
channel, which is equal to one; as $\snr$ goes to infinity,
$C_p(\snr)$ should also converge to one since flash signaling can
be used if there is no peak power constraint (i.e., $\snr=\infty$)
\cite{V02}. Moreover, (\ref{TSNR}) provides a precise
characterization of the interplay between the coherence time and
signal peakiness, stating that the capacity per unit energy is
unaffected as long as the product of $T$ and $\snr$ is fixed. See
\cite{ZTM05,RMZ05} for a related discussion.

For the special case where the fading is a block Gauss-Markov
process, i.e., all the entries of $R(i)$ are equal to $\rho^{i}$
if $i\geq 0$ and equal to $(\rho^*)^{-i}$ if $i<0$ for some
$\rho\in\mathbb{C}$ with $0\leq |\rho|<1$, we have
\begin{eqnarray*}
1+T\snr s(e^{j\omega})=|\varphi(e^{j\omega})|
\end{eqnarray*}
where
\begin{eqnarray*}
\varphi(z)=\frac{\left(\rho^*z-\gamma_0\right)^2}{\gamma_0|\rho|^2\left(z-\frac{1}{\rho^*}\right)^2}
\end{eqnarray*}
and
\begin{eqnarray*}
\gamma_0=\frac{b+\sqrt{b^2-4|\rho|^2}}{2}
\end{eqnarray*}
with $b=1+T\snr+|\rho|^2(1-T\snr)$. The function $\varphi$ is
analytic and nonzero in a neighborhood of the unit disk. Thus, by
Jensen's formula
\begin{eqnarray*}
C_p(\snr)&=&1-\frac{1}{2\pi T\snr}\int_{-\pi}^{\pi}\log|\varphi(e^{j\omega})|\mbox{d}\omega\\
&=&1-\frac{1}{T\snr}\log|\varphi(0)|\\
&=&1-\frac{1}{T\snr}\log \gamma_0,
\end{eqnarray*}
from which we can recover \cite[Corollary 4.1]{SH05} by setting
$T=1$.
\end{example}

Finding the optimal $\mathcal{M}^*$ is a difficult problem in
general. Moreover, as shown in the following example, the optimal
$\mathcal{M}^*$ may depend on the SNR level.

\begin{example}\label{example}
Let the fading process be independent from block to block with
\begin{eqnarray*}
S(e^{j\omega})=R(0)=\begin{pmatrix}
  1 & 1 & \rho^* \\
  1 & 1 & \rho^* \\
  \rho & \rho & 1
\end{pmatrix}
\end{eqnarray*}
where $|\rho|\in[0,1]$.

It is shown in Appendix \ref{app:snrdependent} that
\begin{enumerate}
\item When $0\leq|\rho|\leq\frac{1}{2}$, the optimal
$\mathcal{M}^*$ is $\{1,2\}$, and
\begin{eqnarray*}
C_p(\snr)=1-\frac{1}{2\pi\snr}\Psi(\{1,2\},\snr)=1-\frac{\log(1+2\snr)}{2\snr};
\end{eqnarray*}
\item When $\frac{1}{2}<|\rho|<1$,
\begin{eqnarray*}
\mathcal{M}^*=\left\{\begin{array}{ll} \{1,2,3\} & \snr<\frac{2|\rho|-1}{2(1-|\rho|)^2}\\
\{1,2\}\mbox{ or } \{1,2,3\}   & \snr=\frac{2|\rho|-1}{2(1-|\rho|)^2}\\
\{1,2\} & \snr>\frac{2|\rho|-1}{2(1-|\rho|)^2},
\end{array}\right.&\\
\end{eqnarray*}
and
\begin{eqnarray*}
C_p(\snr)=\left\{\begin{array}{ll} 1-\frac{\log(1+3\snr+2\snr^2-2|\rho|^2\snr^2)}{3\snr} & \snr<\frac{2|\rho|-1}{2(1-|\rho|)^2}\\
1-\frac{\log(1+2\snr)}{2\snr} &
\snr\geq\frac{2|\rho|-1}{2(1-|\rho|)^2};
\end{array}\right.&\\
\end{eqnarray*}
\item When $|\rho|=1$, the optimal $\mathcal{M}^*$ is $\{1,2,3\}$,
and
\begin{eqnarray*}
C_p(\snr)=1-\frac{1}{2\pi\snr}\Psi(\{1,2,3\},\snr)=1-\frac{\log(1+3\snr)}{3\snr}.
\end{eqnarray*}
\end{enumerate}

It can be verified that this result is consistent with the
asymptotic analysis in Corollary \ref{cor:snrasym}. Since
$S_{\mathcal{M}}(e^{j\omega})=R_{\mathcal{M}}(0)$ for any
$\mathcal{M}\subseteq\{1,\cdots,T\}$, it is easy to see that
\begin{eqnarray*}
\sum\limits_{i=0}^{|\mathcal{M}|}\frac{i}{|\mathcal{M}|}\mu(\mbox{rank}(S_{\mathcal{M}}(e^{j\omega}))=i)
\end{eqnarray*}
is minimized at $\mathcal{M}=\{1,2\}$ if $|\rho|<1$, and minimized
at $\mathcal{M}=\{1,2,3\}$ if $|\rho|=1$. Therefore, by
(\ref{highsnrasym}), the optimal $\mathcal{M}^*$ at high SNR
should be $\{1,2\}$ if $|\rho|<1$, and should be $\{1,2,3\}$ if
$|\rho|=1$. Since
\begin{eqnarray*}
&&S^2_{\{1\}}(e^{j\omega})=S^2_{\{2\}}(e^{j\omega})=S^2_{\{3\}}(e^{j\omega})=1,\\
&&S^2_{\{1,3\}}(e^{j\omega})=S^2_{\{2,3\}}(e^{j\omega})=\begin{pmatrix}
  1+|\rho|^2 & 2\rho^* \\
  2\rho & 1+|\rho|^2
\end{pmatrix},\\
&&S^2_{\{1,2\}}(e^{j\omega})=\begin{pmatrix}
  2 & 2 \\
  2 & 2
\end{pmatrix},\\
&&S^2_{\{1,2,3\}}(e^{j\omega})=\begin{pmatrix}
  2+|\rho|^2 & 2+|\rho|^2 & 3\rho^* \\
  2+|\rho|^2 & 2+|\rho|^2 & 3\rho^2 \\
  3\rho & 3\rho & 1+2|\rho|^2
\end{pmatrix},
\end{eqnarray*}
it follows that
$\frac{1}{|\mathcal{M}|}\mbox{tr}\left[S^2_{\mathcal{M}}(e^{j\omega})\right]$
is maximized at $\mathcal{M}=\{1,2\}$ if $|\rho|<\frac{1}{2}$, and
maximized at $\mathcal{M}=\{1,2,3\}$ if $|\rho|>\frac{1}{2}$.
Therefore, by (\ref{lowsnrasym}), the optimal $\mathcal{M}^*$ at
low SNR should be $\{1,2\}$ if $|\rho|<\frac{1}{2}$, and should be
$\{1,2,3\}$ if $|\rho|>\frac{1}{2}$.

Intuitively, if $|\rho|$ is close to 1, we can approximate $R(0)$
by the all-one matrix, and then it follows from Example
\ref{ex:blockindep} that the optimal $\mathcal{M}^*$ is
$\{1,2,3\}$. The approximation breaks down at high SNR since
Corollary \ref{cor:snrasym} implies that the optimal
$\mathcal{M}^*$ should be $\{1,2\}$ as $\snr\rightarrow\infty$.
\end{example}

\section{Conclusion} \label{sec:concl}

We conducted a detailed study of the block-stationary Gaussian
fading channel model introduced in \cite{LV04}. We derived
single-letter upper and lower bounds on channel capacity,
and used these bounds to characterize the asymptotic behavior of
channel capacity. Specifically, we computed the asymptotic ratio
between the non-coherent channel capacity and the logarithm of the
SNR in the high SNR regime. This result generalizes many previous
results on noncoherent capacity. We showed that the behavior of
channel capacity depends critically on channel modelling. We also
derived an expression for the capacity per unit energy for the
block-stationary fading model. It is clearly of interest to
generalize these results to the multi-antenna scenario, but such
an extension seems technically nontrivial.

Another direction that we explored was the interplay between the codeword
length, SNR level, and decoding error probability. We showed that
for noncoherent symbol-by-symbol stationary fading channels, the
codeword length must scale with SNR in order to guarantee that the
communication rate can grow linearly with $\log\snr$ with bounded
decoding error probability, and we found a necessary condition for
the growth rate of the codeword length. We believe that a more complete
characterization of the interplay between the codeword length, SNR
level, and decoding error probability is of both theoretical
significance and practical value.

%We end with some comments on channel modelling. Most existing
%information-theoretic results for non-coherent fading channels,
%including the results given in this paper, require a stationary
%and ergodic model for the channel. However, wireless channels may
%not be well-modelled as stationary processes on a large timescale.
%It might hence be more useful to study outage capacity
%formulations or error exponents for wireless channels.
%However, it appears that existing techniques are not sufficient for
%handling realistic fading channel models.

\appendices
\section{Proof of Monotonicity}\label{app:monotonicity}
By the block stationarity of the fading process, we have
\begin{eqnarray}
I\left(\xb_{i+jT};\yb_{i+jT}\left|\mathbb{E}\left(\hb_{i+jT}\left|\left\{\hb_k+\frac{1}{x_{\min}}\zb_k\right\}_{k=1}^{i+jT-1}\right.\right)\right.\right)=I\left(\xb_{i};\yb_{i}\left|\mathbb{E}\left(\hb_{i}\left|\left\{\hb_k+\frac{1}{x_{\min}}\zb_k\right\}_{k=1-jT}^{i-1}\right.\right)\right.\right).
\label{mono1}
\end{eqnarray}
Since for any $j_1<j_2$,
\begin{eqnarray*}
\xb_{i}\rightarrow\left(\yb_{i},\mathbb{E}\left(\hb_{i}\left|\left\{\hb_k+\frac{1}{x_{\min}}\zb_k\right\}_{k=1-j_2T}^{i-1}\right.\right)\right)\rightarrow\left(\yb_{i},\mathbb{E}\left(\hb_{i}\left|\left\{\hb_k+\frac{1}{x_{\min}}\zb_k\right\}_{k=1-j_1T}^{i-1}\right.\right)\right)
\end{eqnarray*}
form a Markov chain, it follows by data processing inequality that
\begin{eqnarray}
I\left(\xb_{i};\yb_{i}\left|\mathbb{E}\left(\hb_{i}\left|\left\{\hb_k+\frac{1}{x_{\min}}\zb_k\right\}_{k=1-j_1T}^{i-1}\right.\right)\right.\right)\leq
I\left(\xb_{i};\yb_{i}\left|\mathbb{E}\left(\hb_{i}\left|\left\{\hb_k+\frac{1}{x_{\min}}\zb_k\right\}_{k=1-j_2T}^{i-1}\right.\right)\right.\right).\label{mono2}
\end{eqnarray}
Equations (\ref{mono1}) and (\ref{mono2}) together imply that
$\left\{I\left(\xb_{i+jT};\yb_{i+jT}\left|\mathbb{E}\left(\hb_{i+jT}\left|\left\{\hb_k+\frac{1}{x_{\min}}\zb_k\right\}_{k=1}^{i+jT-1}\right.\right)\right.\right)\right\}_j$
is a monotone increasing sequence.

For every
$\mathbb{E}\left(\hb_{i}\left|\left\{\hb_k+\frac{1}{x_{\min}}\zb_k\right\}_{k=1-jT}^{i-1}\right.\right)$,
we can construct a random variable
$\mathbf{\Delta}_j\sim\mathcal{CN}(0,\delta_j)$ independent of
everything else such that
\begin{eqnarray*}
\mathbb{E}\left(\hb_{i}\left|\left\{\hb_k+\frac{1}{x_{\min}}\zb_k\right\}_{k=1-jT}^{i-1}\right.\right)=\mathbb{E}\left(\hb_i\left|\mathbb{E}\left(\hb_{i}\left|\left\{\hb_k+\frac{1}{x_{\min}}\zb_k\right\}_{k=-\infty}^{i-1}\right.\right)+\Delta_j\right.\right)
\end{eqnarray*}
in distribution. Clearly, $\delta_j\rightarrow 0$ as
$j\rightarrow\infty$. Moreover, it is not difficult to show that
\begin{eqnarray}
I\left(\xb_{i};\yb_{i}\left|\mathbb{E}\left(\hb_{i}\left|\left\{\hb_k+\frac{1}{x_{\min}}\zb_k\right\}_{k=1-jT}^{i-1}\right.\right)\right.\right)=I\left(\xb_{i};\yb_{i}\left|\mathbb{E}\left(\hb_{i}\left|\left\{\hb_k+\frac{1}{x_{\min}}\zb_k\right\}_{k=-\infty}^{i-1}\right.\right)+\Delta_j\right.\right).
\label{addnoiseequiv}
\end{eqnarray}
Combining (\ref{mono1}) and (\ref{addnoiseequiv}), we get
\begin{eqnarray*}
&&\lim\limits_{j\rightarrow\infty}I\left(\xb_{i+jT};\yb_{i+jT}\left|\mathbb{E}\left(\hb_{i+jT}\left|\left\{\hb_k+\frac{1}{x_{\min}}\zb_k\right\}_{k=1}^{i+jT-1}\right.\right)\right.\right)\\
&=&\lim\limits_{j\rightarrow\infty}I\left(\xb_{i};\yb_{i}\left|\mathbb{E}\left(\hb_{i}\left|\left\{\hb_k+\frac{1}{x_{\min}}\zb_k\right\}_{k=1-jT}^{i-1}\right.\right)\right.\right)\\
&=&\lim\limits_{j\rightarrow\infty}I\left(\xb_{i};\yb_{i}\left|\mathbb{E}\left(\hb_{i}\left|\left\{\hb_k+\frac{1}{x_{\min}}\zb_k\right\}_{k=-\infty}^{i-1}\right.\right)+\Delta_j\right.\right).
\end{eqnarray*}

Since
\begin{eqnarray*}
&&I\left(\xb_{i};\yb_{i}\left|\mathbb{E}\left(\hb_{i}\left|\left\{\hb_k+\frac{1}{x_{\min}}\zb_k\right\}_{k=-\infty}^{i-1}\right.\right)+\Delta_j\right.\right)\\
&=&I\left(\xb_{i};\yb_{i},\mathbb{E}\left(\hb_{i}\left|\left\{\hb_k+\frac{1}{x_{\min}}\zb_k\right\}_{k=-\infty}^{i-1}\right.\right)+\Delta_j\right)\\
&=&h\left(\yb_{i},\mathbb{E}\left(\hb_{i}\left|\left\{\hb_k+\frac{1}{x_{\min}}\zb_k\right\}_{k=-\infty}^{i-1}\right.\right)+\Delta_j\right)-h\left(\left.\yb_{i},\mathbb{E}\left(\hb_{i}\left|\left\{\hb_k+\frac{1}{x_{\min}}\zb_k\right\}_{k=-\infty}^{i-1}\right.\right)+\Delta_j\right|\xb_{i}\right).
\end{eqnarray*}
By \cite[Lemma 6.11]{LM03}, we get
\begin{eqnarray*}
\lim\limits_{j\rightarrow\infty}h\left(\yb_{i},\mathbb{E}\left(\hb_{i}\left|\left\{\hb_k+\frac{1}{x_{\min}}\zb_k\right\}_{k=-\infty}^{i-1}\right.\right)+\Delta_j\right)=h\left(\yb_{i},\mathbb{E}\left(\hb_{i}\left|\left\{\hb_k+\frac{1}{x_{\min}}\zb_k\right\}_{k=-\infty}^{i-1}\right.\right)\right).
\end{eqnarray*}
Since conditioned on $\xb_{i}$,
$\left(\yb_{i},\mathbb{E}\left(\hb_{i}\left|\left\{\hb_k+\frac{1}{x_{\min}}\zb_k\right\}_{k=-\infty}^{i-1}\right.\right)+\Delta_j\right)$
are jointly Gaussian with uniformly bounded differential entropy
for any realization of $\xb_{i}$ (Note: $|\xb_{i}|^2\leq\snr$), it
follows by dominated convergence theorem that
\begin{eqnarray*}
\lim\limits_{j\rightarrow\infty}h\left(\left.\yb_{i},\mathbb{E}\left(\hb_{i}\left|\left\{\hb_k+\frac{1}{x_{\min}}\zb_k\right\}_{k=-\infty}^{i-1}\right.\right)+\Delta_j\right|\xb_{i}\right)=h\left(\left.\yb_{i},\mathbb{E}\left(\hb_{i}\left|\left\{\hb_k+\frac{1}{x_{\min}}\zb_k\right\}_{k=-\infty}^{i-1}\right.\right)\right|\xb_{i}\right).
\end{eqnarray*}
Therefore,
\[
\lim\limits_{j\rightarrow\infty}I\left(\xb_{i+jT};\yb_{i+jT}\left|\mathbb{E}\left(\hb_{i+jT}\left|\left\{\hb_k+\frac{1}{x_{\min}}\zb_k\right\}_{k=1}^{i+jT-1}\right.\right)\right.\right)\\
=\lim\limits_{j\rightarrow\infty}I\left(\xb_{i};\yb_{i}\left|\mathbb{E}\left(\hb_{i}\left|\left\{\hb_k+\frac{1}{x_{\min}}\zb_k\right\}_{k=-\infty}^{i-1}\right.\right)\right.\right).
\]

\section{Proof of Lemma \ref{lemma:rank}}\label{app:rank}
By eigenvalue decomposition, we write
\begin{eqnarray*}
A(\xi)=U(\xi)D(\xi)U^\dagger(\xi),
\end{eqnarray*}
and
\begin{eqnarray*}
A(\xi)+\epsilon I_M=U(\xi)(D(\xi)+\epsilon I_M)U^\dagger(\xi)
\end{eqnarray*}
where $U(\xi)$ is a unitary matrix, and $D(\xi)$ is a diagonal
matrix with nonnegative diagonal entries. Since
$\mbox{rank}(A(\xi))=\mbox{rank}(D(\xi))$, define
\begin{eqnarray*}
\Omega_i=\{\xi:\mbox{rank}(A(\xi))=\mbox{rank}(D(\xi))=i\},\quad
0\leq i\leq M.
\end{eqnarray*}
We have
\begin{eqnarray*}
\lim\limits_{\epsilon\rightarrow 0
}\frac{\int_{\xi_0}^{\xi_1}\log\det\left[A(\xi)+\epsilon
I_M\right]\mbox{d}\xi}{\log\epsilon}&=&\lim\limits_{\epsilon\rightarrow
0}\frac{\int_{\xi_1}^{\xi_2}\log\det
\left[D(\xi)+\epsilon I_M\right]\mbox{d} \xi}{\log\epsilon}\\
&=&\lim\limits_{\epsilon\rightarrow
0}\frac{-\sum\limits_{i=0}^M\int_{\Omega_i}\log\det
\left[D(\xi)+\epsilon I_M\right]\mbox{d} \xi}{\log\epsilon}.
\end{eqnarray*}
For $\xi\in\Omega_i$, (possibly after permutating
diagonal entries) we can write
$D(\xi)=\mbox{diag}\{d_1(\xi),\cdots,d_i(\xi),0,\cdots,0\}$, where
$d_j(\xi)>0$, $1\leq j\leq i$. Therefore,
\begin{eqnarray*}
\lim\limits_{\epsilon\rightarrow 0
}\frac{\int_{\xi_0}^{\xi_1}\log\det\left[A(\xi)+\epsilon
I_M\right]\mbox{d}\xi}{\log\epsilon}
=\lim\limits_{\epsilon\rightarrow
0}\frac{-\sum\limits_{i=0}^M\int_{\Omega_i}\sum\limits_{j=1}^i\log[d_j(\xi)+\epsilon]\mbox{d}
\xi}{\log\epsilon}+\sum\limits_{i=0}^M(M-i)\mu(\mbox{rank}(A(\xi))=i)
\end{eqnarray*}
where $\mu(\mbox{rank}(A(\xi))=i)$ is the Lebesgue measure of
$\Omega_i$. By the argument in \cite[Section VIII]{L05}, it can be
shown that
\begin{eqnarray*}
\lim\limits_{\epsilon\rightarrow
0}\frac{\int_{\Omega_i}\log[d_i(\xi)+\epsilon]\mbox{d}\xi}{\log\epsilon}=0.
\end{eqnarray*}
So we have
\begin{eqnarray*}
\lim\limits_{\epsilon\rightarrow 0
}\frac{\int_{\xi_0}^{\xi_1}\log\det\left[A(\xi)+\epsilon
I\right]\mbox{d}\xi}{\log\epsilon}
=\sum\limits_{i=0}^M(M-i)\mu(\mbox{rank}(A(\xi))=i).
\end{eqnarray*}

\section{Proof of Theorem \ref{th:asym_cap}}\label{app:asym_cap}

Define
\begin{eqnarray*}
\sigma_i(\snr)=\mbox{var}\left(\hb_i+\frac{1}{\sqrt{\snr}}\zb_i\left|\left\{\hb_k+\frac{1}{\sqrt{\snr}}\zb_k\right\}_{k=-\infty}^{i-1}\right.\right)\quad
i=1,2,\cdots,T.
\end{eqnarray*}
In the lower bound (\ref{lowerbound}), let $\xb_i$ be uniformly
distributed over the set
$\left\{z\in\mathbb{C}:\frac{\sqrt{\snr}}{2}\leq\|z\|\leq\sqrt{\snr}\right\}$.
By Lemma \ref{lemma:computation},
\begin{eqnarray*}
&&I\left(\xb_i;\hb_i\xb_i+\zb_i\left|\mathbb{E}\left(\hb_i\left|\left\{\hb_k+\frac{4}{\snr}\zb_k\right\}_{k=-\infty}^{i-1}\right.\right)\right.\right)\\
&\geq&-\log\left[\sigma_i\left(\frac{\snr}{4}\right)-\frac{12}{5\snr}\right]+\log\left(1-\sigma_i\left(\frac{\snr}{4}\right)+\frac{4}{\snr}\right)-\gamma-\log\frac{5e}{6}\\
&=&-\log\left[\sigma_i\left(\frac{\snr}{4}\right)-\frac{12}{5\snr}\right]+o(\log\snr).
\end{eqnarray*}
Since $\sigma_i\left(\frac{\snr}{4}\right)\geq\frac{4}{\snr}$, it
follows that
\begin{eqnarray*}
\liminf\limits_{\snr\rightarrow\infty}\frac{I\left(\xb_i;\hb_i\xb_i+\zb_i\left|\mathbb{E}\left(\hb_i\left|\left\{\hb_k+\frac{4}{\snr}\zb_k\right\}_{k=-\infty}^{i-1}\right.\right)\right.\right)}{\log\snr}\geq\liminf\limits_{\snr\rightarrow\infty}\frac{-\log\left[\sigma_i\left(\frac{\snr}{4}\right)\right]}{\log\snr}.
\end{eqnarray*}
Let
$\Sigma\left(\frac{\snr}{4}\right)=L\left(\frac{\snr}{4}\right)\Lambda\left(\frac{\snr}{4}\right)L^\dagger\left(\frac{\snr}{4}\right)$,
where $L(\frac{\snr}{4})$ is a lower triangular matrix with unit
diagonal entries, and
$\Lambda\left(\frac{\snr}{4}\right)=\mbox{diag}\left\{\sigma_1\left(\frac{\snr}{4}\right),\sigma_2\left(\frac{\snr}{4}\right),\cdots,\sigma_T\left(\frac{\snr}{4}\right)\right\}$.
We have
\begin{eqnarray}
\det\left[\Sigma\left(\frac{\snr}{4}\right)\right]&=&\det\left[L\left(\frac{\snr}{4}\right)\right]\det\left[\Lambda\left(\frac{\snr}{4}\right)\right]\det\left[L^\dagger\left(\frac{\snr}{4}\right)\right]\nonumber\\
&=&\det\left[\Lambda\left(\frac{\snr}{4}\right)\right]\nonumber\\
&=&\prod\limits_{i=1}^T
\sigma_i\left(\frac{\snr}{4}\right).\label{LDL}
\end{eqnarray}
Therefore,
\begin{eqnarray}
\liminf\limits_{\snr\rightarrow\infty}\frac{C(\snr)}{\log\snr}&\geq&\liminf\limits_{\snr\rightarrow\infty}\frac{\sum\limits_{i=1}^TI\left(\xb_i;\hb_i\xb_i+\zb_i\left|\mathbb{E}\left(\hb_i\left|\left\{\hb_k+\frac{4}{\snr}\zb_k\right\}_{k=-\infty}^{i-1}\right.\right)\right.\right)}{T\log\snr}\nonumber\\
&\geq&\liminf\limits_{\snr\rightarrow\infty}\frac{-\log\left[\prod\limits_{i=1}^T\sigma_i\left(\frac{\snr}{4}\right)\right]}{\log\snr}\nonumber\\
&=&\liminf\limits_{\snr\rightarrow\infty}\frac{-\log\det\left[\Sigma\left(\frac{\snr}{4}\right)\right]}{\log\snr}\nonumber\\
&=&\liminf\limits_{\snr\rightarrow\infty}\frac{-\log\det\left[\Sigma\left(\snr\right)\right]}{\log\snr}.\label{preloginf}
\end{eqnarray}

We use (\ref{upperbound}) derive an upper bound on $\frac{\log
C(\snr)}{\log\snr}$. First it is easy to see that
\begin{eqnarray}
&&\sup\limits_{P_{\xb_k}\in\mathcal{P}_1(\snr)}I\left(\xb_k,\mathbb{E}\left(\hb_k\left|\left\{\hb_t+\frac{1}{\sqrt{\snr}}\zb_{t}\right\}_{t=-\infty}^{k-1}\right.\right);\yb_k\right)\nonumber\\
&\leq&\sup\limits_{P_{\xb_k}\in\mathcal{P}_1(\snr)}I\left(\left.\mathbb{E}\left(\hb_k\left|\left\{\hb_t+\frac{1}{\sqrt{\snr}}\zb_{t}\right\}_{t=-\infty}^{k-1}\right.\right);\yb_k\right|\xb_k\right)+\sup\limits_{P_{\xb_k}\in\mathcal{P}_1(\snr)}I(\xb_k;\yb_k)\label{twoparts}.
\end{eqnarray}
It is shown in \cite{LM03} that
\begin{eqnarray*}
\sup\limits_{P_{\xb_k}\in\mathcal{P}_1(\snr)}I(\xb_k;\yb_k)=o(\log\snr).
\end{eqnarray*}
Now we proceed to upper-bound the first term in (\ref{twoparts}).
\begin{eqnarray*}
&&\sup\limits_{P_{\xb_k}\in\mathcal{P}_1(\snr)}I\left(\left.\mathbb{E}\left(\hb_k\left|\left\{\hb_t+\frac{1}{\sqrt{\snr}}\zb_{t}\right\}_{t=-\infty}^{k-1}\right.\right);\yb_k\right|\xb_k\right)\nonumber\\
&=&\sup\limits_{P_{\xb_k}\in\mathcal{P}_1(\snr)}\mathbb{E}\left\{\log\left[\frac{1+|\xb_k|^2}{1+|\xb_k|^2\cdot\mbox{var}\left(\hb_k\left|\left\{\hb_t+\frac{1}{\sqrt{\snr}}\zb_t\right\}_{t=-\infty}^{k-1}\right.\right)}\right]\right\}\\
&\leq&\log\left[\frac{1+\snr}{1+\snr\cdot\mbox{var}\left(\hb_k\left|\left\{\hb_t+\frac{1}{\sqrt{\snr}}\zb_t\right\}_{t=-\infty}^{k-1}\right.\right)}\right]\\%%+I\left(\left.\hb_{0}+\frac{1}{\sqrt{\snr}}\zb_{0},\cdots,\hb_{2-k}+\frac{1}{\sqrt{\snr}}\zb_{2-k};\yb_1\right|\hb_1,\xb_1\right)\\
&=&\log\left[\frac{1+\snr}{\snr\cdot\sigma_k(\snr)}\right].
\end{eqnarray*}
Therefore,
\begin{eqnarray}
\limsup\limits_{\snr\rightarrow\infty}\frac{C(\snr)}{\log\snr}&\leq&\limsup\limits_{\snr\rightarrow\infty}\frac{\sum\limits_{k=1}^T\sup\limits_{P_{\xb_k}\in\mathcal{P}_1(\snr)}I\left(\left.\mathbb{E}\left(\hb_k\left|\left\{\hb_t+\frac{1}{\sqrt{\snr}}\zb_{t}\right\}_{t=-\infty}^{k-1}\right.\right);\yb_k\right|\xb_k\right)}{T\log\snr}\nonumber\\
&\leq&\limsup\limits_{\snr\rightarrow\infty}\frac{-\log\left[\prod\limits_{k=1}^T\sigma_k\left(\snr\right)\right]}{\log\snr}\nonumber\\
&=&\limsup\limits_{\snr\rightarrow\infty}\frac{-\log\det\left[\Sigma\left(\snr\right)\right]}{\log\snr}.\label{prelogsup}
\end{eqnarray}
The desired result follows by combining (\ref{preloginf}) and
(\ref{prelogsup}).

\section{Proof of Theorem \ref{th:asym_cap2}}\label{app:asym_cap2}
Define
\begin{eqnarray*}
\sigma_i(\infty)=\mbox{var}\left(\hb_i\left|\left\{\hb_k\right\}_{k=-\infty}^{i-1}\right.\right)\quad
i=1,2,\cdots,T.
\end{eqnarray*}
Similar to (\ref{LDL}), we have
\begin{eqnarray*}
\det\left[\Sigma(\infty)\right]=\prod\limits_{i=1}^T\sigma_i(\infty).
\end{eqnarray*}
Therefore, $\det\left[\Sigma(\infty)\right]>0$ implies
$\sigma_i(\infty)>0$ for all $i$.

Note that if $x_{\min}>\delta$ for some $\delta>0$, then
\begin{eqnarray*}
\xb_i\rightarrow\left(\yb_i,\mathbb{E}\left(\hb_{i}\left|\left\{\hb_k+\frac{1}{x_{\min}}\zb_k\right\}_{k=-\infty}^{i-1}\right.\right)\right)\rightarrow
\left(\yb_i,\mathbb{E}\left(\hb_{i}\left|\left\{\hb_k+\delta\zb_k\right\}_{k=-\infty}^{i-1}\right.\right)\right)
\end{eqnarray*}
form a Markov chain, and
\begin{eqnarray*}
I\left(\xb_{i};\hb_i\xb_i+\zb_i\left|\mathbb{E}\left(\hb_{i}\left|\left\{\hb_k+\frac{1}{x_{\min}}\zb_k\right\}_{k=-\infty}^{i-1}\right.\right)\right.\right)\geq
I\left(\xb_{i};\hb_i\xb_i+\zb_i\left|\mathbb{E}\left(\hb_{i}\left|\left\{\hb_k+\delta\zb_k\right\}_{k=-\infty}^{i-1}\right.\right)\right.\right).
\end{eqnarray*}
In the lower bound (\ref{lowerbound}), let $\log|\xb_i|^2$ be
uniformly distributed over the interval $[\log
x^2_{\min},\log\snr]$. As $\log x^2_{\min}$ grows sublinearly in
$\log\snr$ to infinity, we get
\begin{eqnarray*}
&&\liminf\limits_{\snr\rightarrow\infty}\left[I\left(\xb_{i};\hb_i\xb_i+\zb_i\left|\mathbb{E}\left(\hb_{i}\left|\left\{\hb_k+\frac{1}{x_{\min}}\zb_k\right\}_{k=-\infty}^{i-1}\right.\right)\right.\right)-\log\log\snr\right]\\
&\geq&\liminf\limits_{\snr\rightarrow\infty}\left[I\left(\xb_{i};\hb_i\xb_i+\zb_i\left|\mathbb{E}\left(\hb_{i}\left|\left\{\hb_k+\delta\zb_k\right\}_{k=-\infty}^{i-1}\right.\right)\right.\right)-\log\log\snr\right]\\
&=&-1-\gamma-\log\mbox{var}\left(\hb_i\left|\left\{\hb_j+\delta\zb_j\right\}_{j=-\infty}^{i-1}\right.\right)
\end{eqnarray*}
where the last equality follows from \cite[Proposition
4.23]{LM03}. Therefore,
\begin{eqnarray}
&&\liminf\limits_{\snr\rightarrow\infty}\left[C(\snr)-\log\log\snr\right]\nonumber\\
&\geq&\liminf\limits_{\snr\rightarrow\infty}\left[\frac{1}{T}\sum\limits_{i=1}^TI\left(\xb_{i};\hb_i\xb_i+\zb_i\left|\mathbb{E}\left(\hb_{i}\left|\left\{\hb_k+\frac{1}{x_{\min}}\zb_k\right\}_{k=-\infty}^{i-1}\right.\right)\right.\right)-\log\log\snr\right]\nonumber\\
&=&-1-\gamma-\frac{1}{T}\sum\limits_{i=1}^T\log\mbox{var}\left(\hb_i\left|\left\{\hb_j+\delta\zb_j\right\}_{j=-\infty}^{i-1}\right.\right).\label{fadingnumber}
\end{eqnarray}
Since (\ref{fadingnumber}) holds for arbitrary positive $\delta$,
it follows that
\begin{eqnarray}
\liminf\limits_{\snr\rightarrow\infty}[C(\snr)-\log\log\snr]&\geq&-1-\gamma-\frac{1}{T}\lim\limits_{\delta\rightarrow
0}\sum\limits_{i=1}^T\log\mbox{var}\left(\hb_i\left|\left\{\hb_j+\delta\zb_j\right\}_{j=-\infty}^{i-1}\right.\right)\nonumber\\
&=&-1-\gamma-\frac{1}{T}\log\det\left[\Sigma(\infty)\right].\label{fadingnumberinf}
\end{eqnarray}

From the upper bound (\ref{upperbound}), we have
\begin{eqnarray}
&&\limsup\limits_{\snr\rightarrow\infty}[C(\snr)-\log\log\snr]\nonumber\\
&\leq&
\limsup\limits_{\snr\rightarrow\infty}\left[\frac{1}{T}\sum\limits_{k=1}^T\sup\limits_{P_{\xb_k}\in\mathcal{P}_1(\snr)}I\left(\xb_k,\mathbb{E}\left(\hb_k\left|\left\{\hb_t+\frac{1}{\sqrt{\snr}}\zb_{t}\right\}_{t=-\infty}^{k-1}\right.\right);\yb_k\right)-\log\log\snr\right]\nonumber\\
&\leq&\limsup\limits_{\snr\rightarrow\infty}\left[\frac{1}{T}\sum\limits_{k=1}^T\sup\limits_{P_{\xb_k}\in\mathcal{P}_1(\snr)}I\left(\xb_k,\mathbb{E}\left(\hb_k\left|\{\hb_j\}_{j=-\infty}^{k-1}\right);\yb_k\right.\right)-\log\log\snr\right]\label{markov}\\
&\leq&\limsup\limits_{\snr\rightarrow\infty}\left[\sup\limits_{P_{\xb_k}\in\mathcal{P}_1(\snr)}I(\xb_1;\yb_1)+\frac{1}{T}\sum\limits_{k=1}^T\sup\limits_{P_{\xb_k}\in\mathcal{P}_1(\snr)}I\left(\left.\mathbb{E}\left(\hb_k\left|\{\hb_j\}_{j=-\infty}^{k-1}\right.\right);\yb_k\right|\xb_k\right)-\log\log\snr\right]\nonumber
\end{eqnarray}
where (\ref{markov}) follows from the fact that
\begin{eqnarray*}
\left(\xb_k,\mathbb{E}\left(\hb_k\left|\left\{\hb_t+\frac{1}{\sqrt{\snr}}\zb_{t}\right\}_{t=-\infty}^{k-1}\right.\right)\right)\rightarrow\left(\xb_k,\mathbb{E}\left(\hb_k\left|\{\hb_j\}_{j=-\infty}^{k-1}\right.\right)\right)\rightarrow\yb_k
\end{eqnarray*}
form a Markov chain.

It was shown in \cite[Corollary 4.19]{LM03} that
\begin{eqnarray*}
\lim\limits_{\snr\rightarrow\infty}\left[\sup\limits_{P_{\xb_k}\in\mathcal{P}_1(\snr)}I(\xb_1;\yb_1)-\log\log\snr\right]=-1-\gamma.
\end{eqnarray*}
Furthermore,
\begin{eqnarray}
I\left(\left.\mathbb{E}\left(\hb_k\left|\{\hb_j\}_{j=-\infty}^{k-1}\right.\right);\yb_k\right|\xb_k\right)&\leq&
I\left(\left.\mathbb{E}\left(\hb_k\left|\{\hb_j\}_{j=-\infty}^{k-1}\right.\right);\hb_k,\yb_k\right|\xb_k\right)\nonumber\\
&=&I\left(\mathbb{E}(\hb_k|\{\hb_j\}_{j=-\infty}^{k-1});\hb_k\right)+I\left(\mathbb{E}(\hb_k|\{\hb_j\}_{j=-\infty}^{k-1});\yb_k|\hb_k,\xb_k\right)\nonumber\\
&=&I\left(\mathbb{E}(\hb_k|\{\hb_j\}_{j=-\infty}^{k-1});\hb_k\right)\label{Markov2}\\
&=&-\log\sigma_k(\infty)\nonumber
\end{eqnarray}
where (\ref{Markov2}) follows from the fact that
$\mathbb{E}\left(\hb_k|\{\hb_j\}_{j=-\infty}^{k-1}\right)\rightarrow(\hb_k,\xb_k)\rightarrow\yb_k$
form a Markov chain. Therefore, we get
\begin{eqnarray}
\limsup\limits_{\snr\rightarrow\infty}[C(\snr)-\log\log\snr]&\leq&-1-\gamma-\frac{1}{T}\sum\limits_{k=1}^T\log\sigma_k(0)\nonumber\\
&=&-1-\gamma-\frac{1}{T}\log\det\left[\Sigma(\infty)\right].\label{fadingnumbersup}
\end{eqnarray}
The desired result follows by combining (\ref{fadingnumberinf})
and (\ref{fadingnumbersup}).

\section{Proof of Proposition \ref{pro:nonuniform}}\label{app:nonuniform}
At high SNR, we have
\begin{eqnarray*}
\left.\int_{-\pi}^{\pi}\log\left[s_\theta(e^{j\omega})+\frac{4}{\snr}\right]\mbox{d}\omega\right|_{\theta=\snr^{r}}
&=&\alpha\log\left(\frac{4}{\snr}\right)+\left(2\pi-\alpha-\snr^{-r}\right)\log\left(\snr^{-r}+\frac{4}{\snr}\right)\\
&&+\snr^{-r}\log\left(\frac{2\pi\snr^{2r}-2\pi\snr^{r}+\alpha\snr^{r}+1}{\snr^{r}}+\frac{4}{\snr}\right)\\
&=&-2\pi\kappa\log\snr+c(r)+o(1)
\end{eqnarray*}
and thus
\begin{eqnarray}
\mbox{var}\left.\left(\hb_{1}\left|\left\{\hb_k+\frac{2}{\sqrt{\snr}}\zb_1\right\}_{k=-\infty}^{0}\right.\right)\right|_{s_{\theta}(e^{j\omega}),\theta=\snr^{r}}=\frac{e^{\frac{c(r)}{2\pi}}}{\snr^\kappa}-\frac{4}{\snr}+o\left(\frac{1}{\snr^\kappa}\right)\label{predict}
\end{eqnarray}
where $\kappa=\frac{\alpha+\min(r,1)(2\pi-\alpha)}{2\pi}$, and
\begin{eqnarray*}
c(r)=\left\{\begin{array}{ll} \alpha\log4 & r\in(0,1)\\
\alpha\log4+(2\pi-\alpha)\log5 & r=1\\
2\pi\log4 & r\in(1,\infty).
\end{array}\right.&\\
\end{eqnarray*}

In the lower bound (\ref{lowerboundsym}), let $\xb_1$ be uniformly
distributed over the set
$\{z\in\mathbb{C}:\frac{\sqrt{\snr}}{2}\leq |z|\leq\sqrt{\snr}\}$.
By Lemma \ref{lemma:computation} and Equation (\ref{predict}), we
get
\begin{eqnarray*}
C(\snr)&\geq&
I\left(\xb_{1};\hb_1\xb_1+\zb_1\left|\mathbb{E}\left(\hb_{1}\left|\left\{\hb_k+\frac{2}{\sqrt{\snr}}\zb_1\right\}_{k=-\infty}^{0}\right.\right)\right.\right)\\
&\geq&-\log\left[\frac{e^{\frac{c(r)}{2\pi}}}{\snr^\kappa}-\frac{4}{\snr}+\frac{8}{5\snr}+o\left(\frac{1}{\snr^\kappa}\right)\right]+\log\left(1-\frac{e^{\frac{c(r)}{2\pi}}}{\snr}+\frac{4}{\snr}-o\left(\frac{1}{\snr\kappa}\right)\right)-\gamma-\log\frac{5e}{6}\\
&=&\kappa\log\snr+o(\log\snr).
\end{eqnarray*}
Therefore,
\begin{eqnarray*}
\liminf\limits_{\snr\rightarrow\infty,\theta=\snr^{r}}\frac{\left.C(\snr)\right|_{s_\theta(e^{j\omega})}}{\log\snr}\geq
\kappa=\frac{\alpha+\min(r,1)(2\pi-\alpha)}{2\pi}.
\end{eqnarray*}

When $r\geq 1$, we have
\begin{eqnarray*}
\liminf\limits_{\snr\rightarrow\infty,\theta=\snr^{r}}\frac{\left.C(\snr)\right|_{s_\theta(e^{j\omega})}}{\log\snr}\geq
1.
\end{eqnarray*}
Since the noncoherent channel capacity with peak power constraint
$|\xb|^2\leq\snr$ is upper bounded by the coherent channel
capacity with average power constraint
$\mathbb{E}|\xb|^2\leq\snr$, it follows that
\begin{eqnarray*}
\left.C(\snr)\right|_{s_\theta(e^{j\omega})}\leq\mathbb{E}_{\hb}\log\left(1+\snr|\hb|^2\right)\quad\mbox{for
all }\theta.
\end{eqnarray*}
where $\hb\sim\mathcal{CN}(0,1)$. Therefore,
\begin{eqnarray*}
\limsup\limits_{\snr\rightarrow\infty,\theta=\snr^{r}}\frac{\left.C(\snr)\right|_{s_\theta(e^{j\omega})}}{\log\snr}\leq
\lim\limits_{\snr\rightarrow\infty}\frac{\mathbb{E}_{\hb}\log\left(1+\snr|\hb|^2\right)}{\log\snr}=1.
\end{eqnarray*}
The proof is complete.

\section{Example \ref{ex:twolevel}}\label{app:twolevel}

In the lower bound (\ref{lowerboundsym}), let $\xb_1$ be uniformly
distributed over the set
$\{z\in\mathbb{C}:\frac{\sqrt{\snr}}{2}\leq |z|\leq\sqrt{\snr}\}$.
By Lemma \ref{lemma:computation}, we get
\begin{eqnarray}
C(\snr)&\geq&
I\left(\xb_{1};\hb_1\xb_1+\zb_1\left|\mathbb{E}\left(\hb_{1}\left|\left\{\hb_k+\frac{2}{\sqrt{\snr}}\zb_1\right\}_{k=-\infty}^{0}\right.\right)\right.\right)\nonumber\\
&\geq&-\log\left(\mbox{var}\left(\hb_{1}\left|\left\{\hb_k+\frac{2}{\sqrt{\snr}}\zb_1\right\}_{k=-\infty}^{0}\right.\right)+\frac{8}{5\snr}\right)\nonumber\\
&&+\log\left(1-\mbox{var}\left(\hb_{1}\left|\left\{\hb_k+\frac{2}{\sqrt{\snr}}\zb_1\right\}_{k=-\infty}^{0}\right.\right)\right)-\gamma-\log\frac{5e}{6}\label{exlower}
\end{eqnarray}
where
\begin{eqnarray*}
&&\mbox{var}\left(\hb_{1}\left|\left\{\hb_k+\frac{2}{\sqrt{\snr}}\zb_1\right\}_{k=-\infty}^{0}\right.\right)\\
&=&\left(\epsilon_1+\frac{4}{\snr}\right)^{\alpha_1}\left(\epsilon_2+\frac{4}{\snr}\right)^{\alpha_2-\alpha_1}\left[\frac{1-\alpha_1\epsilon_1-(\alpha_2-\alpha_1)\epsilon_2}{1-\alpha_2}+\frac{4}{\snr}\right]^{1-\alpha_2}-\frac{4}{\snr}.
\end{eqnarray*}

By specializing the upper-bound (\ref{upperbound}) to the case
where $T=1$, we obtain
\begin{eqnarray}
C(\snr)&\leq&\sup\limits_{P_{\xb_1}\in\mathcal{P}_1(\snr)}I\left(\xb_1,\mathbb{E}\left(\hb_1\left|\left\{\hb_t+\frac{1}{\sqrt{\snr}}\zb_{t}\right\}_{t=-\infty}^{0}\right.\right);\yb_1\right)\nonumber\\
&\leq&\sup\limits_{P_{\xb_1}\in\mathcal{P}_1(\snr)}I(\xb_1;\yb_1)+\sup\limits_{P_{\xb_1}\in\mathcal{P}_1(\snr)}I\left(\left.\mathbb{E}\left(\hb_1\left|\left\{\hb_t+\frac{1}{\sqrt{\snr}}\zb_{t}\right\}_{t=-\infty}^{0}\right.\right);\yb_1\right|\xb_1\right)\nonumber\\
&\leq&\sup\limits_{P_{\xb_1}\in\mathcal{P}_1(\snr)}I(\xb_1;\yb_1)+\sup\limits_{P_{\xb_1}\in\mathcal{P}_1(\snr)}I\left(\left.\mathbb{E}\left(\hb_1\left|\left\{\hb_t+\frac{1}{\sqrt{\snr}}\zb_{t}\right\}_{t=-\infty}^{0}\right.\right);\hb_1,\yb_1\right|\xb_1\right)\nonumber\\
&=&\sup\limits_{P_{\xb_1}\in\mathcal{P}_1(\snr)}I(\xb_1;\yb_1)+I\left(\mathbb{E}\left(\hb_1\left|\left\{\hb_t+\frac{1}{\sqrt{\snr}}\zb_{t}\right\}_{t=-\infty}^{0}\right.\right);\hb_1\right)\label{exupper}
\end{eqnarray}
where
\begin{eqnarray*}
&&I\left(\mathbb{E}\left(\hb_1\left|\left\{\hb_t+\frac{1}{\sqrt{\snr}}\zb_{t}\right\}_{t=-\infty}^{0}\right.\right);\hb_1\right)\\
&=&-\log\left[\mbox{var}\left(\hb_1\left|\left\{\hb_t+\frac{1}{\sqrt{\snr}}\zb_{t}\right\}_{t=-\infty}^{0}\right.\right)\right]\\
&=&-\log\left\{\left(\epsilon_1+\frac{1}{\snr}\right)^{\alpha_1}\left(\epsilon_2+\frac{1}{\snr}\right)^{\alpha_2-\alpha_1}\left[\frac{1-\alpha_1\epsilon_1-(\alpha_2-\alpha_1)\epsilon_2}{1-\alpha_2}+\frac{1}{\snr}\right]^{1-\alpha_2}-\frac{1}{\snr}\right\}.
\end{eqnarray*}
Note that $\sup_{P_{\xb_1}\in\mathcal{P}_1(\snr)}I(\xb_1;\yb_1)$
is the capacity of the memoryless noncoherent Rayleigh fading
channel (see \cite{LM03} for a nonasymptotic upper bound), and we
have
$\sup_{P_{\xb_1}\in\mathcal{P}_1(\snr)}I(\xb_1;\yb_1)\ll\log\snr$
for large $\snr$.

By (\ref{exlower}) and (\ref{exupper}), it is not hard to verify
that $\frac{C(\snr)}{\log\snr}$ is approximately equal to
$\alpha_2$ for $1\ll\snr\leq\frac{1}{\epsilon_2}$, and is
approximately equal to $\alpha_1$ for
$\log\frac{1}{\epsilon_2}\ll\log\snr\leq\log\frac{1}{\epsilon_1}$.

\section{AWGN Channel}\label{app:AWGN}

By the random coding bound \cite{G68}, we have
\begin{eqnarray*}
P_e\leq e^{-nE_r(R)}
\end{eqnarray*}
where
\begin{eqnarray}
E_r(R)=\frac{\snr}{2e^R}\left[e^R+1-(e^R-1)\sqrt{1+\frac{4e^R}{\snr(e^R-1)}}\right]+\log\left\{e^R-\frac{\snr(e^R-1)}{2}\left[\sqrt{1+\frac{4e^R}{\snr(e^R-1)}}-1\right]\right\}\label{exponent1}
\end{eqnarray}
if
$\log\left[\frac{1}{2}+\frac{\snr}{4}+\frac{1}{2}\sqrt{1+\frac{\snr^2}{4}}\right]\leq
R\leq\log(1+\snr)$, and
\begin{eqnarray}
E_r(R)=1+\frac{\snr}{2}-\sqrt{1+\frac{\snr^2}{4}}+\log\left(\frac{1}{2}-\frac{\snr}{4}+\frac{1}{2}\sqrt{1+\frac{\snr^2}{4}}\right)+\log\left(\frac{1}{2}+\frac{\snr}{4}+\frac{1}{2}\sqrt{1+\frac{\snr^2}{4}}\right)-R\label{exponent2}
\end{eqnarray}
if
$R<\log\left[\frac{1}{2}+\frac{\snr}{4}+\frac{1}{2}\sqrt{1+\frac{\snr^2}{4}}\right]$.

Let $R(\snr)=\log\snr-\log\eta$ where $\eta\in(1,2)$. By
(\ref{exponent1}),
\begin{eqnarray*}
&&\lim\limits_{\snr\rightarrow\infty}E_r(R(\snr))\\
&=&\lim\limits_{\snr\rightarrow\infty}\frac{\eta}{2}\left[\frac{\snr}{\eta}+1-\left(\frac{\snr}{\eta}-1\right)\sqrt{1+\frac{4}{\snr-\eta}}\right]+\log\left\{\frac{\snr}{\eta}-\frac{\snr(\snr-\eta)}{2\eta}\left[\sqrt{1+\frac{4}{\snr-\eta}}-1\right]\right\}\\
&=&\lim\limits_{\snr\rightarrow\infty}\frac{\eta}{2}\left[\frac{\snr}{\eta}+1-\frac{\snr-\eta}{\eta}\left(1+\frac{2}{\snr-\eta}\right)\right]+\log\left\{\frac{\snr}{\eta}-\frac{\snr(\snr-\eta)}{2\eta}\left[\frac{2}{\snr-\eta}-\frac{2}{(\snr-\eta)^2}\right]\right\}\\
&=&\eta-1-\log\eta\\
&>0&.
\end{eqnarray*}
For any $P_e>0$, we can find an $n$ such that
\begin{eqnarray*}
e^{-n\left(\eta-1-\log\eta\right)}<P_e.
\end{eqnarray*}
Therefore, for any $P_e>0$, there exist a sequence of codebooks
with rate $R(\snr)$ and fixed codeword length $n$ such that
\begin{eqnarray*}
\lim\limits_{\snr\rightarrow\infty}\frac{R(\snr)}{\log\snr}=1
\end{eqnarray*}
and $\limsup_{\snr\rightarrow\infty}P_e(\snr)\leq P_e$.

%If $R(\snr)=\alpha\log\snr$ where $\alpha\in(0,1)$, by
%(\ref{exponent2})
%\begin{eqnarray*}
%E_r(R(\snr))=(1-\alpha)\log\snr+o(\log\snr)
%\end{eqnarray*}
%at high $\snr$. In this case, we can let $n=1$, and still get
%\begin{eqnarray*}
%\lim\limits_{\snr\rightarrow\infty}\frac{R(\snr)}{\log\snr}=\alpha
%\end{eqnarray*}
%and $\limsup_{\snr\rightarrow\infty}P_e(\snr)\leq P_e$ for any
%$P_e>0$.

\section{Coherent Rayleigh Fading Channel}\label{app:Rayleigh}

It was shown in \cite{T99} that
\begin{eqnarray*}
E_r(R)=\max\limits_{0\leq\rho\leq
1}\left[-\log\mathbb{E}_{\hb}\left(1+\frac{\snr}{1+\rho}|\hb|^2\right)^{-\rho}-\rho
R\right]
\end{eqnarray*}
where $\hb\sim\mathcal{CN}(0,1)$.

Choosing $R(\snr)=\log\snr-\log\log\snr-c$ and $\rho=1$, we get
\begin{eqnarray*}
\liminf\limits_{\snr\rightarrow\infty}E_r(R(\snr))&\geq&
\liminf\limits_{\snr\rightarrow\infty}\left[-\log\mathbb{E}_{\hb}\left(1+\frac{\snr}{1+\rho}|\hb|^2\right)^{-1}-\log\snr+\log\log\snr+c\right]\\
&=&\liminf\limits_{\snr\rightarrow\infty}\left[-\log\mathbb{E}_{\hb}\left(\frac{1}{\snr}+\frac{1}{2}|\hb|^2\right)^{-1}+\log\log\snr+c\right]\\
&=&\liminf\limits_{\snr\rightarrow\infty}\left\{-\log\left[\int_{0}^\infty\left(\frac{1}{\snr}+\frac{t}{2}\right)^{-1}e^{-t}\mbox{d}t\right]+\log\log\snr+c\right\}\\
&=&\liminf\limits_{\snr\rightarrow\infty}\left\{-\log\left[\int_{0}^1\left(\frac{1}{\snr}+\frac{t}{2}\right)^{-1}e^{-t}\mbox{d}t+\int_{1}^\infty\left(\frac{1}{\snr}+\frac{t}{2}\right)^{-1}e^{-t}\mbox{d}t\right]+\log\log\snr+c\right\}\\
&\geq&\liminf\limits_{\snr\rightarrow\infty}\left\{-\log\left[\int_{0}^1\left(\frac{1}{\snr}+\frac{t}{2}\right)^{-1}\mbox{d}t+\int_{1}^\infty2e^{-t}\mbox{d}t\right]+\log\log\snr+c\right\}\\
&=&\liminf\limits_{\snr\rightarrow\infty}\left[-\log\left[2\log(2+2\snr+\snr^2)-2\log(2+2\snr)+2e^{-1}\right]+\log\log\snr+c\right]\\
&=&-\log 2+c
\end{eqnarray*}
which is positive if $c>\log 2$.

Therefore, for any $P_e>0$, we can find a sequence of codebooks
with rate $R(\snr)$ and fixed codeword length $n$ such that
\begin{eqnarray*}
\lim\limits_{\snr\rightarrow\infty}\frac{R(\snr)}{\log\snr}=1
\end{eqnarray*}
and $\limsup_{\snr\rightarrow\infty}P_e(\snr)\leq P_e$.

\section{Example \ref{example}}\label{app:snrdependent}

We can compute that
\begin{eqnarray*}
&&\Psi(\{1\},\snr)=\Psi(\{2\},\snr)=\Psi(\{3\},\snr)=2\pi\log(1+\snr),\\
&&\Psi(\{1,3\},\snr)=\Psi(\{2,3\},\snr)=\pi\log(1+2\snr+\snr^2-|\rho|^2\snr),\\
&&\Psi(\{1,2\},\snr)=\pi\log(1+2\snr),\\
&&\Psi(\{1,2,3\},\snr)=\frac{2\pi}{3}\log(1+3\snr+2\snr^2-2|\rho|^2\snr^2).
\end{eqnarray*}
It can be verified that
\begin{eqnarray*}
\Psi(\{1\},\snr)=\Psi(\{2\},\snr)=\Psi(\{3\},\snr)\geq\Psi(\{1,3\},\snr)=\Psi(\{2,3\},\snr)\geq\Psi(\{1,2\},\snr).
\end{eqnarray*}
So the optimal $\mathcal{M}^*$ is either $\{1,2\}$ or $\{1,2,3\}$.
Setting $\Psi(\{1,2\},\snr)=\Psi(\{1,2,3\},\snr)$ yields
\begin{eqnarray*}
(1+3\snr+2\snr^2-2|\rho|^2\snr^2)^2=(1+2\snr)^3
\end{eqnarray*}
which, after some algebraic manipulation, is equivalent to
\begin{eqnarray*}
1-4|\rho|^2+(4-12|\rho|^2)\snr+(4-8|\rho|^2+4|\rho|^4)\snr^2=0.
\end{eqnarray*}
The above equation has two solutions
\begin{eqnarray*}
\snr_1=\frac{-2|\rho|-1}{2(1+|\rho|)^2},\quad\snr_2=\frac{2|\rho|-1}{2(1-|\rho|)^2}.
\end{eqnarray*}
$\snr_1$ can be discarded since it is always negative. $\snr_2$ is
positive for $|\rho|\in(\frac{1}{2},1)$. When
$|\rho|\in(\frac{1}{2},1)$, it can be verified that
$\Psi(\{1,2\},\snr)>\Psi(\{1,2,3\},\snr)$ if $\snr<\snr_2$, and
$\Psi(\{1,2\},\snr)<\Psi(\{1,2,3\},\snr)$ if $\snr>\snr_2$. When
$|\rho|\in[0,\frac{1}{2}]$, $\snr_2$ is non-positive. In this
case, we have $\Psi(\{1,2\},\snr)>\Psi(\{1,2,3\},\snr)$ for all
$\snr>0$.

\end{document}